\documentclass{emulateapj}
\usepackage{amsmath,amssymb,threeparttable}

\shorttitle{$\beta$ Pictoris}
\begin{document}

\title{A SMACK Model of Colliding Planetesimals in the $\beta$ Pictoris Debris Disk}

\author{Erika R. Nesvold}
\affil{Department of Physics, University of Maryland Baltimore County 
\\ 1000 Hilltop Circle
\\ Baltimore, MD 21250}
\affil{Department of Terrestrial Magnetism, Carnegie Institution for Science
\\ 5241 Broad Branch Rd
\\ Washington, DC 20015}
\email{enesvold@carnegiescience.edu}

\author{Marc J. Kuchner}
\affil{NASA Goddard Space Flight Center 
\\ Exoplanets and Stellar Astrophysics Laboratory, Code 667
\\ Greenbelt, MD 21230}
\email{Marc.Kuchner@nasa.gov}

\begin{abstract}

We present a new model of the $\beta$ Pictoris disk-and-planet system that simulates both the planetesimal collisions and the dynamics of the resulting dust grains, allowing us to model features and asymmetries in both thermal and scattered light images of the disk. Our two-part model first simulates the collisional and dynamical evolution of the planetesimals with the Superparticle-Method Algorithm for Collisions in Kuiper belts (SMACK) and then simulates the dynamical evolution of the resulting dust grains with a standard Bulirsch-Stoer N-body integrator. Given the observed inclination and eccentricity of the $\beta$ Pictoris b planet, the model neatly ties together several features of the disk: the central hole in the submillimeter images, the two-disk ``x''-pattern seen in scattered light, and possibly even the clumpy gas seen by ALMA. We also find that most of the dust in the $\beta$ Pictoris system is likely produced outside the ring at 60-100 AU. Instead of a birth ring, this disk has a ``stirring ring'' at 60-100 AU where the high-velocity collisions produced by the secular wave launched by the planet are concentrated. The two-disk x-pattern arises because collisions occur more frequently at the peaks and troughs of the secular wave. The perturbations of the disk in this region create an azimuthally and vertically asymmetric spatial distribution of collisions, which could yield an azimuthal clump of gas without invoking resonances or an additional planet.

\end{abstract}

\section{Introduction}
\label{sec:introduction}

The $\beta$ Pictoris system (a 21 Myr, A6 star at 19.44 pc) harbors a debris disk whose angular size, brightness, and intriguing morphology make it one of the most-observed disks in the sky. The disk was first detected by \citet{Aumann1984} with the Infrared Astronomical Satellite (IRAS). \citet{Smith1984} took the first resolved image of the disk using a coronagraph on the du Pont 2.5-m telescope at the Las Campanas Observatory, revealing a nearly edge-on system. Eleven years after the first resolved image, more detailed images revealed asymmetries \citep{Kalas1995}, including a pattern they referred to as a ``warp'' in the inner region of the disk \citep{Burrows1995, Heap2000}. In high-resolution scattered light images of the disk, the warp resembles two distinct, inclined disks, forming a x-shaped pattern in the images \citep{Golimowski2006, Ahmic2009, Apai2015}. \citet{Mouillet1997} and later \citet{Augereau2001} were able to simulate a warp (but not the ``x''-pattern) in numerical simulations by including a planet with an orbit inclined to the main disk. 

\citet{Lagrange2010} later discovered a planet, $\beta$ Pictoris b, orbiting between 8 and 15 AU from the star, and estimated its mass as $9\pm3$ $M_{Jup}$, in agreement with the model predictions of \citet{Mouillet1997} and \citet{Augereau2001}. Initial measurements of the planet's orbit indicated that it might be misaligned with the second inclined disk \citep{Currie2011}, but modeling indicated that, within the observational uncertainties, the observed planet could still be sufficiently inclined to the main disk to be responsible for the warp \citep{Dawson2011}. More recent observations have been able to constrain the orbit of $\beta$ Pic b and confirm that the planet is inclined to the main disk \citep{Chauvin2012, Nielsen2014}. 

Submillimeter observations of the disk show a broad belt of planetesimals orbiting at $94\pm8$ AU \citep{Wilner2011, Dent2014}, orbiting on the same plane as the extended main disk of smaller dust grains observed in scattered light \citep[][etc.]{Golimowski2006}. Submillimeter and infrared images of the $\beta$ Pic disk have revealed a variety of additional asymmetries in the $\beta$ Pictoris disk. \citet{Wahhaj2003} and \citet{Telesco2005} observed a bright clump of emission in the mid-infrared in the SW region of the disk at a projected separation of $\sim50$ AU. \citet{Li2012} confirmed this mid-infrared clump at 52  AU, but noted a spatial displacement between the two epochs, which they attributed to Keplerian motion. The nearly edge-on viewing geometry of the $\beta$ Pictoris debris disk complicates attempts to unravel the 3D structure of the disk due to degeneracies along the line-of-sight. Using ALMA, \citet{Dent2014} observed a clump of short-lived CO gas, also in the SW region, but orbiting with a true orbital distance of $\sim85$ AU from the star, indicating an azimuthally-asymmetric region of enhanced collisions. \citet{Apai2015} compared HST/STIS observations of the disk from epochs 15 years apart and were unable to detect Keplerian motion from any point source contributing $> 3 \%$ of the disk surface brightness at projected separations between $3\farcs0$ and $5\farcs0$ (58 and 97 AU). However, this region of the disk does not directly probe the CO or mid-infrared clumps.

Interpreting these patterns in the $\beta$ Pic disk and in other debris disks requires modeling collisions between planetesimals in addition to the dynamics of the system \citep{Stark2009, Thebault2012, Thebault2012a, Charnoz2012, Nesvold2013, Kral2013, Nesvold2015, Kral2015}. In this paper we use the term ``planetesimal'' to refer to the parent bodies in the disk larger than 1 mm in size.The collisional lifetime of a $\sim10$ cm parent body orbiting in the $\beta$ Pic disk at 10 AU is $\sim2\times10^6$ yr (assuming a low optical depth of $10^{-4}$), while at 100 AU, the collision time is $\sim6\times10^7$ yr. This timescale is only three times the age of the system, 21 Myr \citep{Binks2014}, and smaller bodies will have even shorter timescales for fragmenting collisions, indicating that collisions are occurring frequently enough to influence the evolution of the disk. Collisions likely play some role in the observed ALMA asymmetry \citep{Dent2014}, highlighting the need for a model incorporating planetesimal collisions. New models of the disk also need to show how the recent best-fit planet orbits from \citet{Nielsen2014} influence the disk morphology.

Interpreting the scattered light images calls for new dynamical models of the dust in the $\beta$ Pic disk \citep{Mouillet1997, Augereau2001, Dawson2011}. While existing models can replicate the ``butterfly asymmetry'' \citep{Kalas1995} or warp induced by the planet, they fail to reproduce the distinct two-disk x-pattern seen in high-resolution scattered light images \citep[e.g., Fig. 9 of][]{Golimowski2006}. The dynamics of the dust differ from the dynamics of the planetesimals, as the dust is subject to radiation pressure and Poynting-Robertson drag \citep{Robertson1937}. The distribution of the dust also depends on the spatial distribution and collision rate of the planetesimals that produce the dust via collisions, as well as the destructive collisions between the dust grains. 

To answer this call, we present simulations of the $\beta$ Pictoris debris disk using our 3D collisional disk model SMACK, which traces the location of dust-producing collisions within a disk simultaneously with the evolving spatial distribution of larger planetesimals ($>1$ mm). We combine these with a dynamical model of the dust to create the first physical model of both the planetesimal collisions and the dust dynamics in this system. In Section \ref{sec:smack} we describe the SMACK simulations and the parameters used. In Section \ref{sec:relaxation} we measure the level of collisional relaxation within the disk to determine whether collisional damping is significant. In Section \ref{sec:spiral} we discuss two kinds of spiral structures created by the slightly-eccentric $\beta$ Pic b, and its implications for observations of asymmetries in the disk. We also discuss the origin of the observed central hole in the distribution of mm-sized planetesimals. In Section \ref{sec:collisions} we discuss the 3D distribution of collisions in the disk. In Section \ref{sec:dust} we present a model of the dust in the $\beta$ Pic disk, created by integrating the orbits of the dust grains produced in SMACK collisions, and in Section \ref{sec:scatteredlight} we present simulated images and a radial brightness profile of the dust in disk, created using our dust models. The model we describe in this paper does not include the fragmentation of grains smaller than 1 mm and is therefore an incomplete model of the small grain population, so in Section \ref{sec:limitations}, we discuss the implications of these limitations. Finally, in Section \ref{sec:conclusions}, we summarize our results and propose future work with this model.

\section{SMACK Simulations of Colliding Planetesimals}
\label{sec:smack}

Modeling the dynamical and collisional evolution of the parent bodies in a disk is essential for accurately calculating the effects of collisional damping and collisional erosion on the morphology of a disk \citep{Nesvold2013, Nesvold2015}. (We do not include a full collisional model of the dust in this work. See Section \ref{sec:limitations} for a discussion of the implications of this omission.) 

Our debris disk simulator, the Superparticle-Method Algorithm for Collisions in Kuiper Belts (SMACK), simulates the evolution of the dynamics and the spatially-dependent size distribution of a disk of planetesimals under the influence of one or more planets. SMACK uses the N-body integrator REBOUND \citep{Rein2012}, but each body in the integrator represents a superparticle, a cloud of planetesimals with the same position and trajectory, but a range of sizes from 1 mm-10 cm. Each superparticle is characterized by a size distribution. When an overlap is detected between two superparticles, SMACK adjusts the size distributions and trajectories of the colliding superparticles to represent the outcome of fragmenting planetesimal collisions. Data output by SMACK include the time-dependent 3D density map of the planetesimals and a 3D map of the dust-producing collisions throughout the simulation time. 

We simulated the evolution of the $\beta$ Pictoris system with SMACK using 100,000 superparticles for a simulation time of 21 Myr, the age of $\beta$ Pic as measured by \citet{Binks2014}. We inserted a planet of mass 9 $M_{Jup}$, using the best-fit orbit of \citet{Nielsen2014}, who used the Gemini/NICI and Magellan/MagAO instruments to measure the position of $\beta$ Pic b relative to its star with greater accuracy than with previous observations alone \citep[e.g.,][]{Chauvin2012, Bonnefoy2013}. We distributed the initial superparticle orbits uniformly and linearly in semi-major axis, eccentricity, inclination, longitude of ascending node, argument of pericenter, and mean anomaly, and assigned their size distributions to be power laws with index $-3.5$, normalized such that the initial optical depth of the disk at 100 AU was $10^{-2}$. Note that a linear distribution in semi-major axis will result in a radial surface density distribution of $r^{-1}$. This radial distribution is typically for gas disk models \citep[see][for a discussion of surface density profiles]{Raymond2014}. The initial parameters of the superparticles and planet are listed in Table \ref{tab:initial}. We selected a superparticle size of $r_{sp} = 10^{-1.5}$ AU. The finite superparticle size limits the size of the features that SMACK can resolve. At the location of the planet, the superparticle size we chose corresponds to an inclination and an eccentricity of $r_{sp}/a \approx 0.18^{\circ}$ and $r_{sp}/a \approx 0.003$, respectively. Note that the superparticle encounter timescale does not represent the collisional timescale in the disk. The SMACK algorithm requires that the superparticle encounter timescale is shorter than the collisional timescale. It also requires a superparticle size small enough that the outcome of the model does not depend on the exact choice of superparticle size \citep{Nesvold2013}. The simulation ran for $\sim270$ hr of wall clock time on the NASA Center for Climate Simulation's Discover cluster, using a hybrid OpenMP/MPI parallelization on 48 cores. 

\begin{deluxetable}{lcc}
\tablewidth{0pt}
\tablecaption{Initial conditions of the superparticles and planet for the simulation described in Section \ref{sec:smack}. \label{tab:initial}}
\tablehead{\colhead{Parameter	} & \colhead{Superparticle Range} & \colhead{Planet Value}}
\startdata
	Semi-Major Axis 	 		& 5-200 					& 9.1 \\
	Eccentricity 				& 0.0-0.01 				& 0.08\\
	Inclination 				& 0.0-0.005 				& 0.0452\\
	Long. of Ascending Node	 	& 0-$2\pi$ 				& 0\\
	Argument of Periapsis 		& 0-$2\pi$ 				& 0.113\\
	Mean Anomaly 				& 0-$2\pi$ 				& 0\\
\enddata
\tablecomments{All angular values are given in radians.}
\end{deluxetable}

Although we ran the simulation for 21 Myr, the most recent estimate for the age of the star \citep{Binks2014}, the planet may have been formed or scattered onto its current inclined orbit more recently than the star was formed \citep{Dawson2011, Currie2013}, so 21 Myr does not necessarily represent the age of the planet-star system in its current configuration. \citet{Mouillet1997} derived a relationship between the radial extent of the warp, $r_w$, and the system age, $t_{age}$,
\begin{equation} \label{eq:speed} r_w = 6.31 \left[ \frac{m_{pl}}{M_{*}} \left(\frac{r_{pl}}{10 \mbox{ AU}}\right)^2 \frac{t_{age}}{5.2 \mbox{ yr}} \right]^{0.29}, \end{equation}
where $m_{pl}$ is the planet's mass, $M_{*}$ is the star's mass, and $r_{pl}$ is the planet's orbital radius. For example, for a system with our parameters, in which we place the planet on its inclined orbit at time $t=0$, the warp should extend out to $\sim86$ AU after 10 Myr, while the maximum extent at 21 Myr would be $\sim107$ AU. Observations of the disk in scattered light and submillimeter emission tend to place the extent of the warp in the planetesimals and dust between $\sim85-95$ AU, so we synthesized images from our simulations after 10 Myr of evolution rather than 21 Myr. 

\begin{figure*} [!ht]
	\centering
	\includegraphics[trim=0 150 0 150,clip,width=0.98\textwidth]{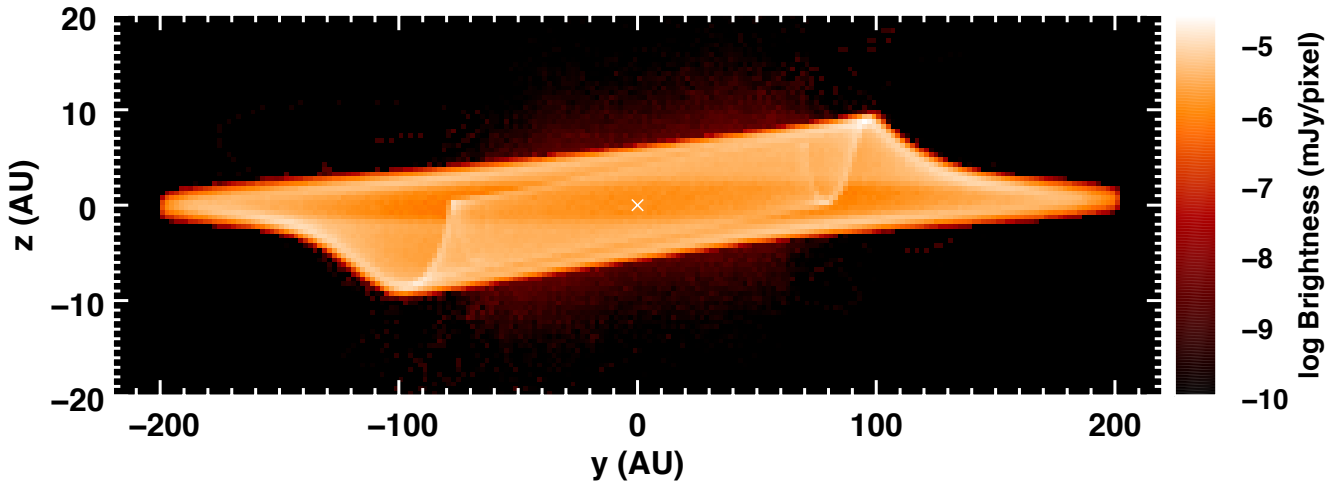}
	\caption{Simulated SMACK image of the $\beta$ Pictoris disk at 850 $\mu$m after 10 Myr. The location of the star is indicated with the white x. The vertical axis has been stretched to emphasize the warp structure. The simulated disk after 10 Myr resembles the collisionless simulation of \citet{Dawson2011}, but in our disk model, collisions have destroyed most of the planetesimals that have completed a full secular oscillation.}
	\label{fig:edgeon}
\end{figure*}

Figure \ref{fig:edgeon} shows a simulated image of the $\beta$ Pictoris disk at 10 Myr at a wavelength of 850 $\mu$m. To simulate the image, we assumed that each planetesimal within a superparticle emits thermally as a spherical blackbody, and iteratively calculated the temperature of each planetesimal as a function of its distance from the star. We summed the contribution from each superparticle in a given pixel, and averaged over 50 output timesteps, or $5\times10^5$ yr, to reduce the Poisson noise in the image. Note that at 50 AU, this is roughly two orders of magnitude less than the secular timescale of $\sim4.7\times10^{6}$ yr, ensuring that features due to secular perturbations will not be smeared. Separate models of the dust dynamics are not needed to compare to ALMA images because the basic SMACK models include all the grain sizes that emit efficiently at millimeter wavelengths. 

Our simulations reproduce the general morphology of previous models of the warp created by the inclined orbit of the planet \citep[e.g.,][]{Augereau2001,Dawson2011}. At 10 Myr after the addition of an inclined planet, the warp has propagated out to $\sim100$ AU. The shape of the disk as simulated with SMACK is similar to the results from collisionless simulations \citep[e.g., Figure 1 in][]{Dawson2011}, except that our simulated image exhibits a deficit of the material represented by orange dots in \citet{Dawson2011}. We discuss this inner clearing in more detail in Section \ref{sec:clearing}. 

\begin{figure*} [!ht]
	\centering
	\includegraphics[trim=0 150 0 150,clip,width=0.98\textwidth]{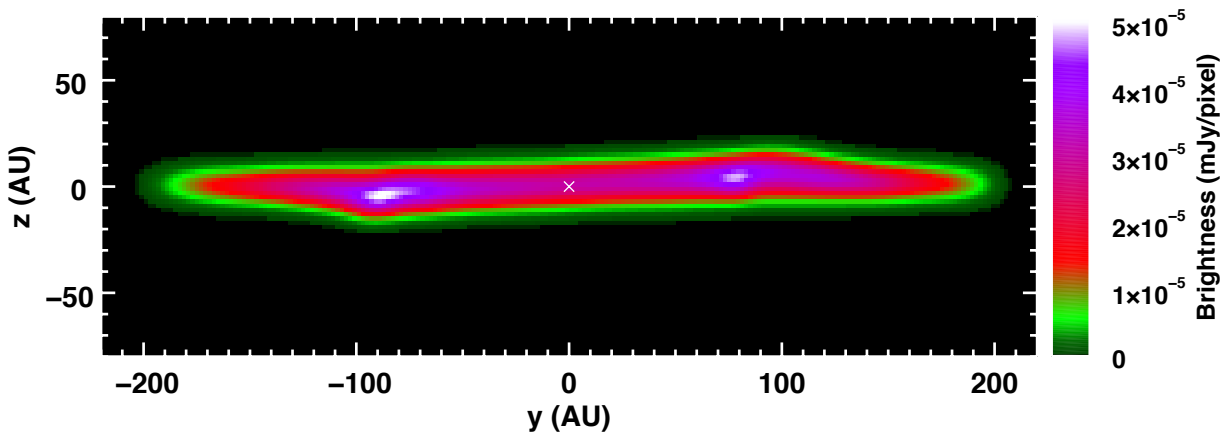}
	\caption{Simulated SMACK image of the $\beta$ Pictoris disk shown in Figure \ref{fig:edgeon}, with linear color scaling and no vertical stretch. This image has been convolved with a Gaussian beam with a FWHM of 12 AU.  Comparing this simulated image with the ALMA continuum observation by \citet{Dent2014}, we find that we reproduce the basic morphology of the ALMA image, but our brightness peaks are located farther out and our simulation does not reproduce the brightness asymmetry.}
	\label{fig:alma}
\end{figure*}

Figure \ref{fig:alma} shows the same simulated image at 850 $\mu$m, with a linear scaling and no vertical stretch. We convolved this image with a Gaussian beam with a full-width half-max (FWHM) of 12 AU to simulate an ALMA observation. Clearly, our simulations reproduce the basic morphology seen in the ALMA image, but there are some consequential differences between the model and observations. Comparing Figure \ref{fig:alma} to the continuum image of the disk with ALMA \citep{Dent2014}, we note that the brightness peaks correspond to the radial extent of the warp, which is $\sim95$ AU at 10 Myr in our simulation \citep[rather than the $\sim86$ AU predicted by][]{Mouillet1997}. In the ALMA observation, the surface brightness peaks at $\sim60$ AU on either side of the star, indicating that $\beta$ Pictoris b reached its current inclined orbit $< 10$ Myr ago, assuming the planet mass is $\geq 9 M_{Jup}$. While the simulated disk exhibits a small brightness asymmetry in this viewing orientation, our simulation is unable to reproduce the $\sim15\%$ brightness asymmetry between the SW and NE halves of the disk seen in the ALMA image (see Section \ref{sec:spiral} for a further discussion of the brightness asymmetry). 

\section{Collisional Relaxation}
\label{sec:relaxation}

SMACK models have the novel ability to numerically explore the proposed process of collisional relaxation in debris disks. This process, the gradual removal of free eccentricity and free inclination from planetesimal orbits, has been invoked in influential models of debris disks with eccentric rings \citep[e.g.,][]{Quillen2007, Chiang2009}, but not investigated numerically until \citet{Nesvold2013}, in which the authors demonstrated the destruction, via collisions, of a spiral density wave created by an eccentric planet in a disk, leaving a narrow eccentric debris ring. 

The morphology of the $\beta$ Pictoris disk, however, indicates that collisions may not have completely damped the free inclinations in the disk, but the collisional damping in $\beta$ Pic has not been studied by previous models. We describe here what our simulations of $\beta$ Pic disk show about this process.

A planet on an orbit inclined to a disk will impose a forced inclination on the disk's planetesimals. A planetesimal's inclination, $i$, and longitude of ascending node, $\Omega$, can be written together as a vector with components
\begin{eqnarray}
p & = & i \sin \Omega \\
q & = & i \cos \Omega.
\end{eqnarray}
The planetesimal's inclination $\bf i$ will precess about the forced inclination $\mathbf{i}_{forced}$ induced by the planet, such that
\begin{equation}  \label{eq:inclinations} \mathbf{i} = \mathbf{i}_{free} + \mathbf{i}_{forced}, \end{equation}
where $\mathbf{i}_{free}$ is the free or proper inclination of the planetesimal \citep{Murray1999}. Inelastic collisions will tend to damp the free inclinations of the superparticles. 

\begin{figure*} [!ht]
	\centering
	\includegraphics[trim=0 120 0 100,clip,width=0.98\textwidth]{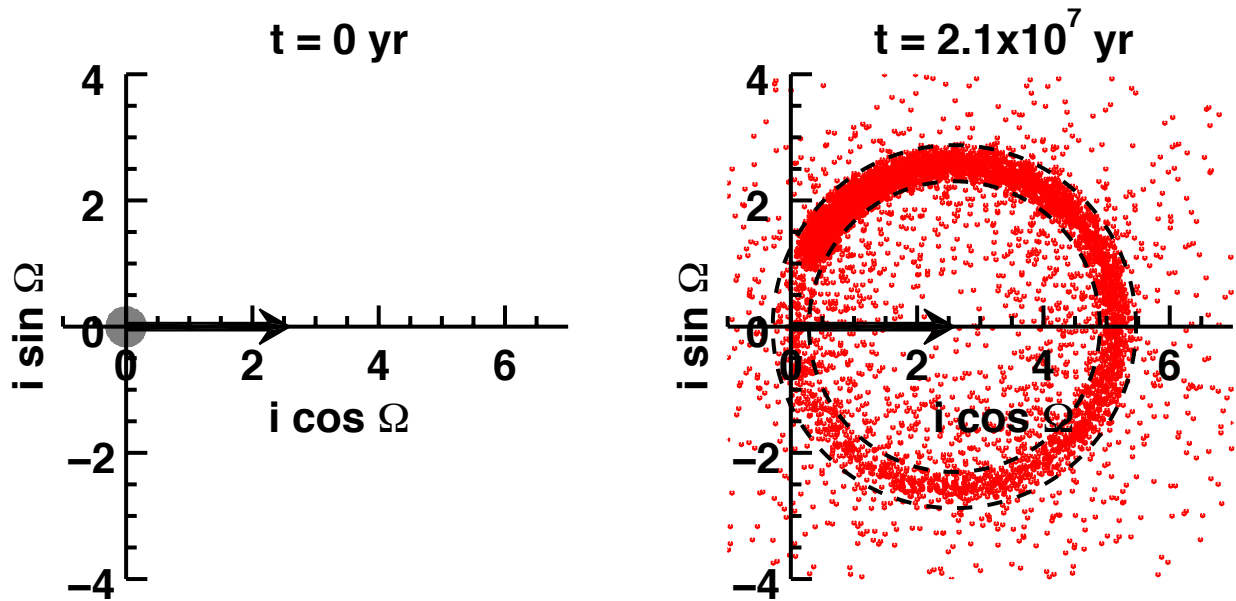}
	\caption{The $p$-$q$ diagrams of all of the superparticles at the beginning and end of the simulation described in Section (\ref{sec:smack}). The black arrow represents the forced inclination vector due to the planet. The dashed black lines illustrate the annulus in which the inclination vectors of the superparticles would precess in the absence of collisions. Ninety-six percent of the mass in the simulation is within the annulus, indicating that minimal collisional damping has occurred by $2.1\times10^{7}$ yr.}
	\label{fig:hkdiagram}
\end{figure*}

Plotting the inclination vectors of the planetesimals in $p$-$q$ space can help illustrate this relationship and process. In Figure \ref{fig:hkdiagram}, we plot the $p$-$q$ diagrams of the superparticles at various times during the simulation. The black arrow represents the forced inclination vector imposed by the planet. At $t=0$, the superparticles have inclinations uniformly distributed in a small range of values, and are uniformly distributed in $\Omega$ from $0-2\pi$. They are represented on the $p$-$q$ diagram as a small grey circle. As the system begins to evolve, the superparticles' inclination precess about the planet's forced inclination at different rates, spreading the superparticles into a ring in $p$-$q$ space. The magnitude of the superparticles' inclination vectors oscillates between their initial inclinations near $\sim0$ and twice the planet's inclination, creating a two-component disk, with a ``secondary'' disk inclined to the main disk by twice the planet's inclination.

In a completely damped system, the superparticles will have zero free inclination, so by Equation (\ref{eq:inclinations}), their inclinations will all equal the forced inclination of the planet, leaving a single disk in the plane of the planet's orbit rather than the two-component disk seen in $\beta$ Pic and with collisionless simulations \citep[e.g.,][]{Dawson2011}. On a $p$-$q$ diagram, the points representing completely damped superparticles would appear clustered at the planet's inclination vector. We can see from the right panel in Figure \ref{fig:hkdiagram} that the superparticles are barely damped at all after 21 Myr. Some superparticles (representing 3\% of the mass of the system) have moved inwards towards the planet's inclination vector due to collisional damping of their free inclinations; others (1\% of the system by mass) have been collisionally scattered into higher inclinations. However, most (96\% of the mass in the simulation) remains within the annulus defined by the range of their initial inclinations, indicating that the gravitational perturbations of the planet dominate the dynamical effects of collisions, and that the system is not yet collisionally damped. This lack of damping is consistent with recent measurements of the orbit of $\beta$ Pic b relative to the disk with Gemini/NICI and Magellan/MagAO, which indicate that the planet's position angle lies in between the position angles of the main disk and the inner warp on the sky \citep{Nielsen2014}. 

\begin{figure*} [!ht]
	\centering
	\includegraphics[trim=0 0 0 0,clip,width=0.98\textwidth]{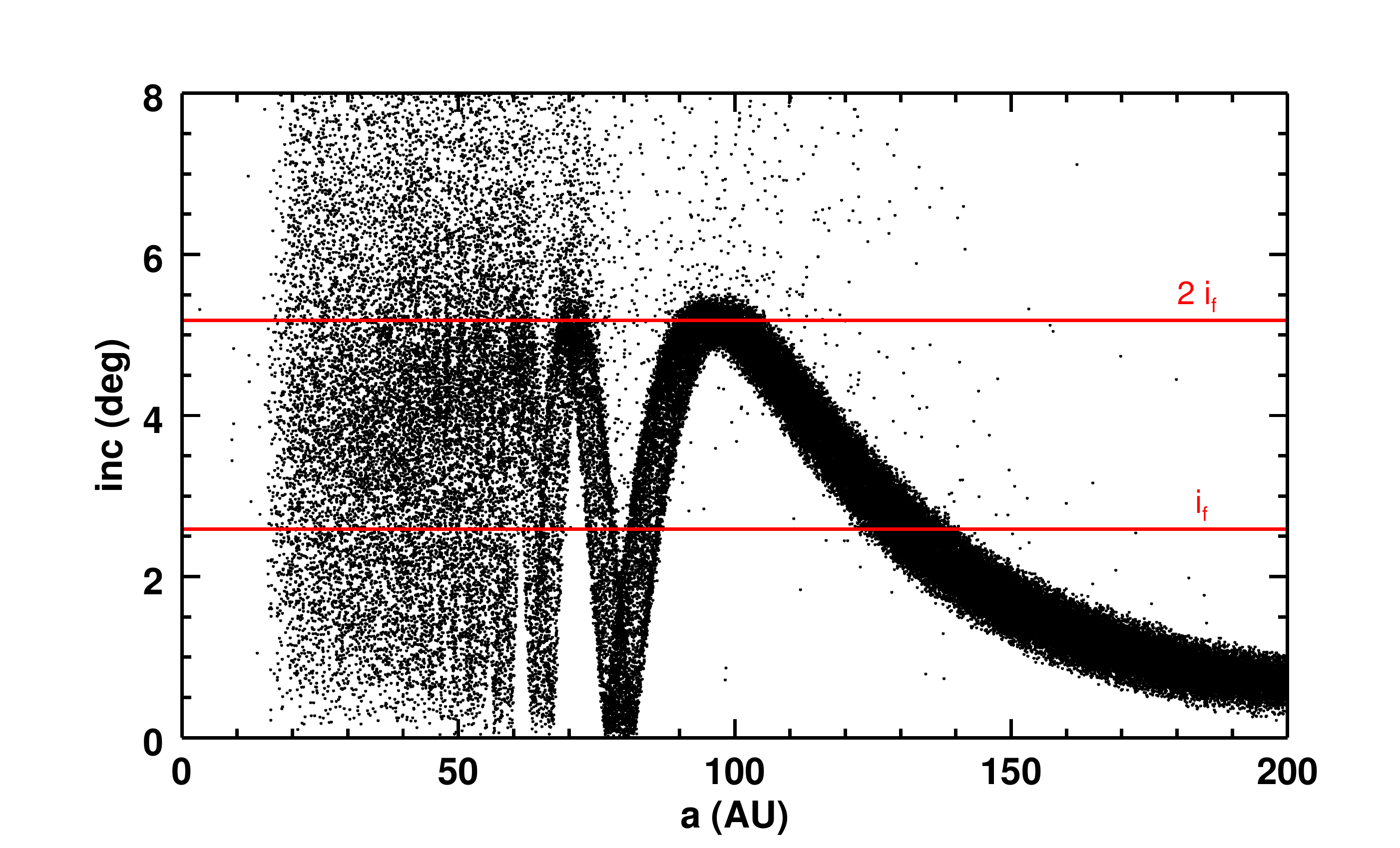}
	\caption{Inclination vs. semi-major axis of the superparticles at 10 Myr. The red lines indicate the planet's forced inclination, $i_{f}$, and $2i_{f}$. This figure strongly resembles the results of \citet{Dawson2011} for $a\gtrsim59$ AU, but at $a<59$ AU in our simulations, collisions dominate and scatter planetesimals to high inclination.}
	\label{fig:ivsa} 
\end{figure*}

Following the example of \citet{Dawson2011}, we plotted the inclination vs. semi-major axis of the superparticles at 10 Myr in Figure \ref{fig:ivsa}. Again, the results are similar to the collisionless results of \citet{Dawson2011}. The forced inclination from the planet, $i_{f}$, is independent of radial distance to the planet and simply equals the planet's inclination, $i_{f}=i_{pl}$. The superparticles orbiting at $\gtrsim170$ AU are still at low inclination, the superparticles within $\sim100-150$ AU have just reached their maximum inclination of $2i_{f}$ for the first time, and the superparticles inside $\sim100$ AU are oscillating between the inclination of the main disk (around 0) and $2i_{f}$. However, in the SMACK models, we see the effects of collisions in the superparticles that have been collisionally scattered to both lower- and higher-inclination ($\>2i_{f}$) orbits. Again, there is little evidence of collisional damping, which would manifest as a clustering of superparticles at the planet's inclination $i_{f}$, beginning with the superparticles farthest in, closest to the planet's orbit.

In a single collision between planetesimals, SMACK decreases of the kinetic energy of the planetesimals by a factor of 0.1 to represent the energy used to fragment the bodies \citep[following][]{Fujiwara1982}. Given our previous estimate that the collisional lifetime of a 10 cm body at 100 AU in the $\beta$ Pictoris disk is 60 Myr, this indicates that random velocities in the disk will decrease by a factor of $1/e$ after 220 Myr, consistent with the lack of discernible damping during our 21 Myr simulation. This estimation assumes that the damping rate is constant, which will not be the case as the damping will decrease the collision rate in the disk. Further simulations are needed to predict the timescale of the collisional damping in $\beta$ Pictoris and the future morphology of the disk.

\section{Spiral Structure}
\label{sec:spiral}

\subsection{Simulation Results}

Just as the planet's inclination can create a warp in a debris disk, secular perturbations from an eccentric planet will perturb the orbits of the particles in the disk. This effect will create a spiral structure in an initially axisymmetric disk, as planetesimals at different distances from the planet precess at different rates. This spiral will spread radially outwards from the planet's orbit. Interior to the spiral, the orbits of the planetesimals are phase mixed, so the spiral structure only persists until the spiral reaches the outer edge of the disk \citep{Wyatt2005a}. Collisionless numerical simulations of the $\beta$ Pictoris disk by \citet{Mouillet1997}, \citet{Matthews2014} and \citet{Apai2015} predict that an eccentric $\beta$ Pic b will induce a spiral in the disk via this mechanism.

As with inclination, the eccentricity of a particle can be written as a vector using the particle's longitude of pericenter, $\varpi$: $\mathbf{e} = (e \sin \varpi, e \cos \varpi)$. Laplace-Lagrange secular theory provides an expression for a planetesimal's eccentricity vector $\mathbf{e}(t)$ perturbed by an eccentric planet. If we assume that the initial eccentricity of each planetesimal is zero, $\mathbf{e}(0)=\mathbf{0}$, the eccentricity of a planetesimal with semi-major axis $a$ is 
\begin{equation} \label{eq:ecc}
\begin{split}
\mathbf{e}(t) = \bigg( \frac{b^2_{3/2}(\alpha_{pl})}{b^1_{3/2}(\alpha_{pl})} e_{pl} \cos \varpi_{pl} (1+\cos At),\\ \frac{b^2_{3/2}(\alpha_{pl})}{b^1_{3/2}(\alpha_{pl})} e_{pl} \sin \varpi_{pl} (1+\sin At) \bigg),
\end{split}
\end{equation}
where $e_{pl}$ is the planet's eccentricity, $\varpi_{pl}$ is the planet's longitude of pericenter, $b^2_{3/2}$ and $b^1_{3/2}$ are Laplace coefficients, and $\alpha_{pl}=a/a_{pl}$ for $a>a_{apl}$, where $a_{pl}$ is the semi-major axis of the planet \citep{Murray1999,Wyatt1999a}. The precession rate $A$ of the planetesimal is given by
\begin{equation} \label{eq:rate} A = \frac{n}{4} \frac{m_{pl}}{M_*} \alpha_{pl} \bar{\alpha}_{pl} b^1_{3/2}(\alpha_{pl}), \end{equation}
where $n$ is the planetesimal's mean motion, $m_{pl}$ is the mass of the planet, $M_*$ is the mass of the star, and $\bar{\alpha}_{pl}=1$ for $a>a_{pl}$ \citep{Murray1999,Wyatt1999a}.  The apsidal precession period is then given by $2\pi/A$. To first order, the planetesimal's inclination vector $\mathbf{i}(t)$, perturbed by an inclined planet, will precess with the same precession rate $A$, and the forced inclination is simply equal to the inclination of the planet, $i_{pl}$:
\begin{equation} \label{eq:inc}
\mathbf{i}(t) = \left( i_{pl} \cos \Omega_{pl} (1+\cos At), i_{pl} \sin \Omega_{pl} (1+\sin At) \right),
\end{equation}
where $\Omega_{pl}$ is the longitude of the ascending node of the planet \citep{Murray1999}.

\begin{figure*} [!ht]
	\centering
	\includegraphics[scale=0.6]{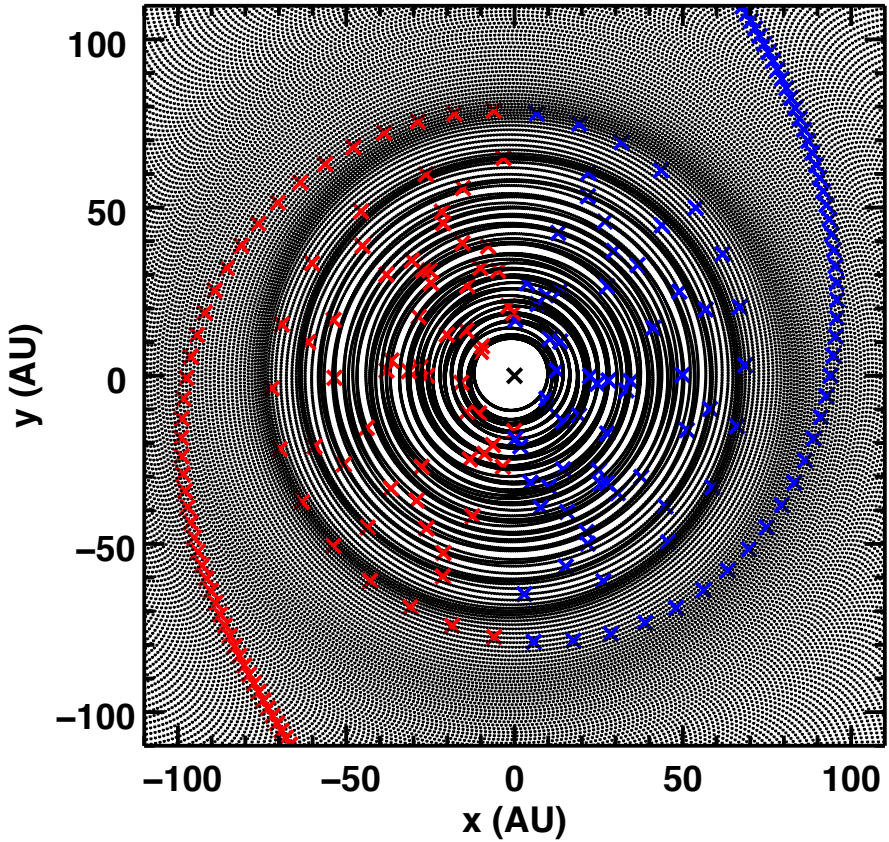}
	\caption{Face-on diagram of the analytically-derived orbits of planetesimals perturbed by a planet with the orbit and mass of $\beta$ Pic b after 10 Myr. The black x marks the location of the star. Orbits at different semi-major axes precess at different rates, forming a spiral density wave where they approach adjacent orbits. Blue and red x's mark the locations where the derived orbits cross the $z=0$ plane, indicating the ascending nodes and descending nodes of each orbit, respectively. They trace a double-armed spiral that intersects the spiral density wave (see Section \ref{sec:collisions}).} 
	\label{fig:orbits}
\end{figure*}

To better understand the geometry of the perturbed disk, we used Equations \ref{eq:ecc}-\ref{eq:inc} to analytically calculate the effects of the forced eccentricity and inclination from $\beta$ Pic b on the orbits after 10 Myr using the planet parameters listed in Table \ref{tab:initial} and plotted the resulting orbits in Figure \ref{fig:orbits}. We considered the orbits of planetesimals on initially circular orbits with semi-major axes ranging from 11 AU to 155 AU. As described in \citet{Wyatt2005a}, the orbits precess at different rates, forming a spiral density wave extending radially outward to $\sim 100$ AU. Interior to $\sim 59$ AU, the planetesimals have completed more than one precession period and their orbits have become phase mixed, while exterior to $\sim 100$ AU, the planetesimals are still orbiting with very low eccentricities. 

As we described in Section \ref{sec:relaxation}, the inclinations of the planetesimals precess around the forced eccentricity in a similar manner: the planetesimals interior to $\sim59$ AU have completed at least one precession period, while planetesimals exterior to $\sim100$ AU still have very low inclinations (Figure \ref{fig:ivsa}). Rather than a spiral density wave, however, the secular effects of the planet's inclination induce a vertical displacement wave in the planetesimals. In Figure \ref{fig:orbits} we mark the ascending and descending nodes of the orbits with blue and red x's, respectively. The nodes form a double-armed spiral, indicating that the vertical displacement wave varies azimuthally. We will refer to the spiral density wave created by the planet's eccentricity as the ``eccentricity spiral'' and the vertical displacement wave created by the planet's inclination as the ``inclination spiral''. 

\begin{figure*} [!ht]
	\centering
	\includegraphics[scale=0.6]{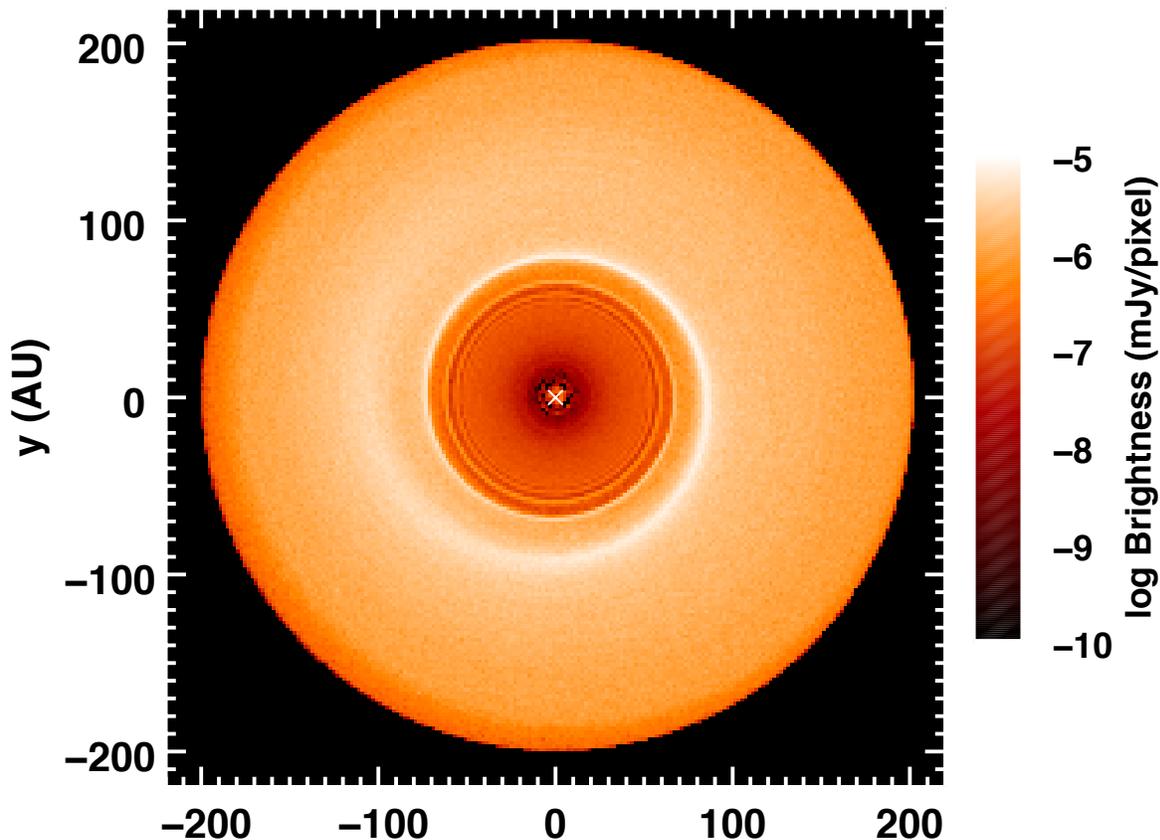}
	\caption{Face-on simulated image of the $\beta$ Pictoris disk at 850 $\mu$m after 10 Myr (seen edge-on in Figure \ref{fig:edgeon}). The white x indicates the location of the star. The disk at $\lesssim60$ AU is roughly an order of magnitude fainter than the disk exterior to the spiral structure because collisions have destroyed the planetesimals in this central clearing.} 
	\label{fig:faceon}
\end{figure*}

Although collisions can eventually destroy an eccentricity spiral in a disk \citep{Nesvold2013}, the eccentricity spiral induced in the $\beta$ Pictoris disk survives to 21 Myr. Figure \ref{fig:faceon} shows a simulated image of the face-on $\beta$ Pictoris disk at 10 Myr, the same simulation shown in Figure \ref{fig:edgeon}. A spiral structure is evident, extending radially outward to $\sim100$ AU, in good agreement with the analytically-predicted spiral shown in Figure \ref{fig:orbits}.

The outermost portion of the spiral in Figure \ref{fig:faceon} corresponds to the planetesimals that have completed half a precession period, and have reached their eccentricity maximum. Since the precession rate for inclinations and eccentricity are equal to first order, these planetesimals have also reached their maximum inclination, so the outermost spiral in the $\beta$ Pic disk is roughly co-located with the maximum radial extent of the warp. 

\citet{Wilner2011} observed the $\beta$ Pictoris disk with the Submillimeter Array and detected two peaks in 1.3 mm emission along the disk plane, which they interpreted as a ring or belt of larger, dust-producing planetesimals at $94\pm8$ AU, with a deficit of mm-sized grains interior to the belt. Our simulation results suggest that if the warp in the $\beta$ Pic disk extends to $\sim85$ AU \citep{Golimowski2006,Heap2000}, a spiral structure created by $\beta$ Pic b's eccentricity will also extend out to $\sim85$ AU. The ``ring'' of planetesimals observed by \citet{Wilner2011} may, in fact, be this spiral structure.

\citet{Apai2015} proposed that the spiral structure could contribute a small brightness asymmetry to the disk. \citet{Wilner2011} noted that the SW side of the disk appeared brighter in their SMA observations, but the difference was below the noise level. \citet{Dent2014} observed the $\beta$ Pic disk at 870 $\mu$m and found that the continuum emission from the SW side of the disk is $15\%$ brighter, on average, than from the NE side. 

To test whether the spiral structure could explain this asymmetry, we calculated the simulated emission from the NE and SW halves of the disk, observed edge-on, while rotating the disk about its axis, at 1 Myr intervals from $t=0$ to $t=10$ Myr. We found that the brightness ratio of the NE and SW regions of the disk varies with the orientation of the spiral at all timesteps. The maximum possible brightness excess of the SW disk versus the NE disk peaks at $\sim18\%$ at 1 Myr, but drops to less than $5\%$ after the disk evolves past 4 Myr. By 10 Myr, the maximum possible brightness asymmetry was only $\sim2\%$. So we infer that the brightness asymmetry due to the spiral may contribute to the observed brightness excess in the SW disk, but probably cannot be solely responsible for it. Note that the spiral structure propagates outwards with time and does not orbit the star, indicating that the brightness asymmetry due to the spiral will not move to the opposite side of the star, but will instead move radially outwards with time. 

\subsection{Central Clearing}
\label{sec:clearing}

The brightness of the disk shown in Figure \ref{fig:faceon} drops off interior to the ring (at radii $\lesssim59$ AU) by roughly an order of magnitude compared to regions of the disk exterior to the spiral. \citet{Wilner2011} observed a deficit of mm-sized grains at radii $\lesssim 94$ AU. \citet{Dent2014} also observed that the mm-sized grains lie in a belt with a central clearing. However, the mechanism for clearing these larger grains from the interior region of the disk is not immediately obvious. The planet will clear a gap around its orbit via resonance overlap, but this gap will only extend out to $\sim14.5$ AU in 10 Myr even accounting for the effects of collisions, which tend to widen the gap \citep{Nesvold2015}.

\begin{figure*} [!ht]
	\centering
	\includegraphics[trim=0 0 0 0,clip,width=0.98\textwidth]{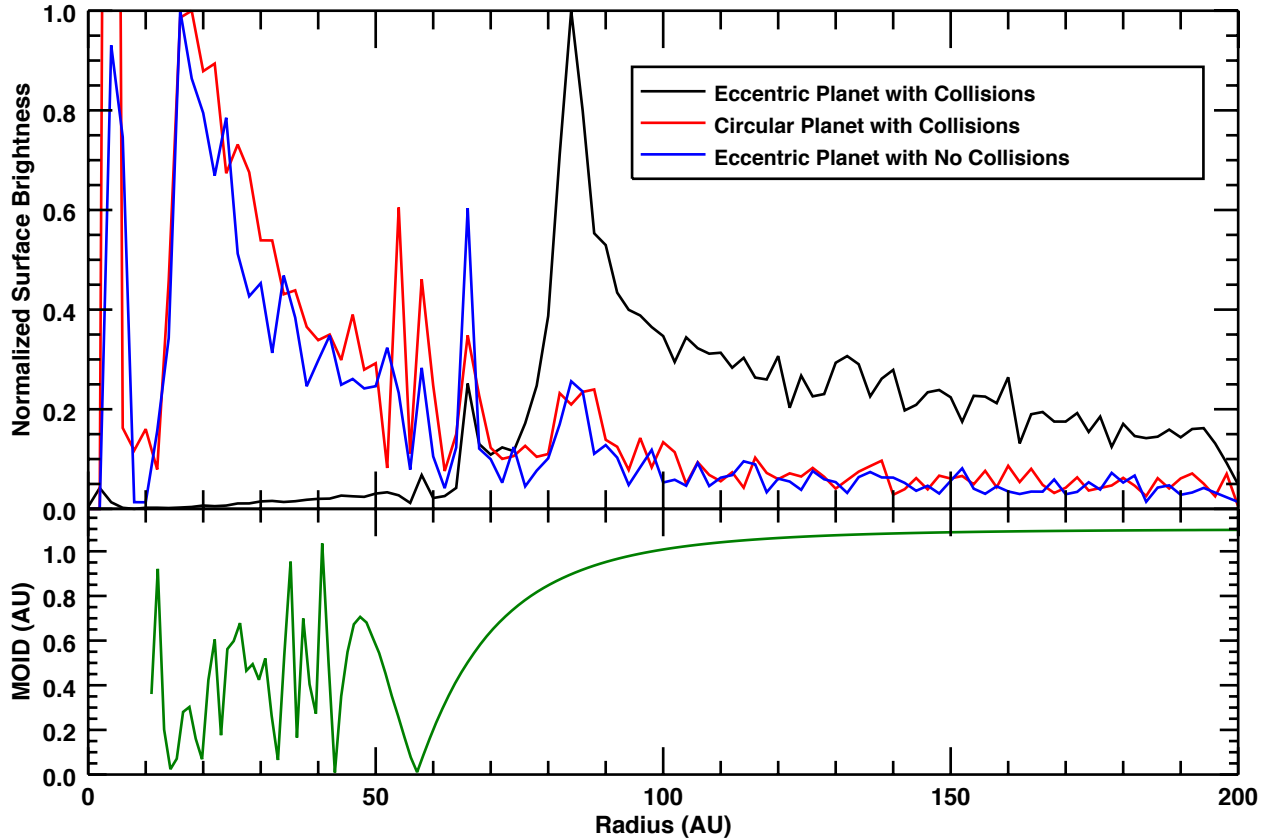}
	\caption{Top panel: Normalized 850 $\mu$m surface brightness vs. radius at 10 Myr for three simulations: the disk with an eccentric planet (shown in Figure \ref{fig:faceon}), the same simulation with the planet's eccentricity set to zero, and the eccentric planet simulation with the collisions turned off. An eccentric planet in a collisional disk (black curve) creates a central brightness deficit in mm-sized bodies. Bottom panel: Minimum orbit intersection distance (MOID) versus semi-major axis for pairs of adjacent orbits in Figure \ref{fig:orbits}. Orbits in the interior region of the disk ($\lesssim59$ AU) intersect, increasing the collision rate and creating the brightness deficit see in the black curve in the top panel.}
	\label{fig:radmoid}
\end{figure*}

Our model suggests that a different mechanism is producing the central clearing of planetesimals in the $\beta$ Pic disk. The top panel of Figure \ref{fig:radmoid} shows the normalized radial surface brightness for the simulated disk shown in Figure \ref{fig:faceon}, compared with two new simulations, each run with 10,000 superparticles for 10 Myr. In one simulation, we set the eccentricity of the planet to zero. In the other, we kept the planet's eccentricity set to $e=0.08$, but we turned off the collision simulation and let the system evolve with only dynamical perturbations. Only the system with an eccentric planet in a disk experiencing collisions shows a brightness deficit interior to $\sim59$ AU. 

\citet{Mustill2009} showed that a planet's secular perturbations will stir a disk of planetesimals and place them on intersecting orbits. Where planetesimal orbits cross, collisions are more frequent, and colliding bodies will be gradually eroded and removed from the system. They derived an expression for $t_{cross}$, the time required for a planet's secular perturbations to cause two neighboring planetesimal orbits to intersect. For the planet and star parameters in our SMACK simulation, $t_{cross} \approx 10$ Myr at a semi-major axis of 60 AU in the disk. This is illustrated further in the bottom panel of Figure \ref{fig:radmoid}, where we plot the minimum orbit intersection distance (MOID) versus semi-major axis for pairs of adjacent orbits from Figure \ref{fig:orbits}, calculated using the method described in \citet{Wisniowski2013}. The last zero of the MOID appears to be at $\sim59$ AU. Exterior to 59 AU, the MOID increases, then asymptotes to a value of 1.1 AU, which was the separation of the orbits of the planetesimals in our analytic calculations. The orbit crossings depicted by the green curve must cause the collisions that create the central clearing depicted by the black curve. The orbit-crossing timescale of \citet{Mustill2009} indicates that the region of orbit crossing (and, therefore, the central clearing) will reach 94 AU at $\sim88$ Myr.

The two methods of disk clearing discussed in this section, resonance overlap \citep{Nesvold2015} and secular excitation \citep{Mustill2009}, invoke two different sets of initial conditions for the disk. The \citet{Nesvold2015} model assumes that the planetesimals in the disk have some initial eccentricity distribution, such that planetesimals orbiting just outside the resonance overlap region of the planet (the planet's ``chaotic zone'') will collide frequently and widen the gap. The \citet{Mustill2009} mechanism assumes that the disk is initially cold, and depends on secular perturbations from the planet to excite collisions between planetesimals. Our simulation results for the $\beta$ Pictoris system suggest that the mechanism described by \citet{Mustill2009} dominates in this system, widening the cleared inner region to much greater radial distances than the resonance clearing mechanism can reach in the age of the system. Future simulations should explore the differences between these ``hot-start'' and ``cold-start'' disk models, and investigate what observed disk clearing can tell us about the initial conditions of the disk.
 
\section{Size and Spatial Distribution of Collisions} 
\label{sec:sizeandspace}
 
 \subsection{Spatial Distribution of Collisions}
 \label{sec:collisions} 
 
\begin{figure*} [!ht]
	\centering
	\includegraphics[trim=0 100 0 90,clip,width=0.98\textwidth]{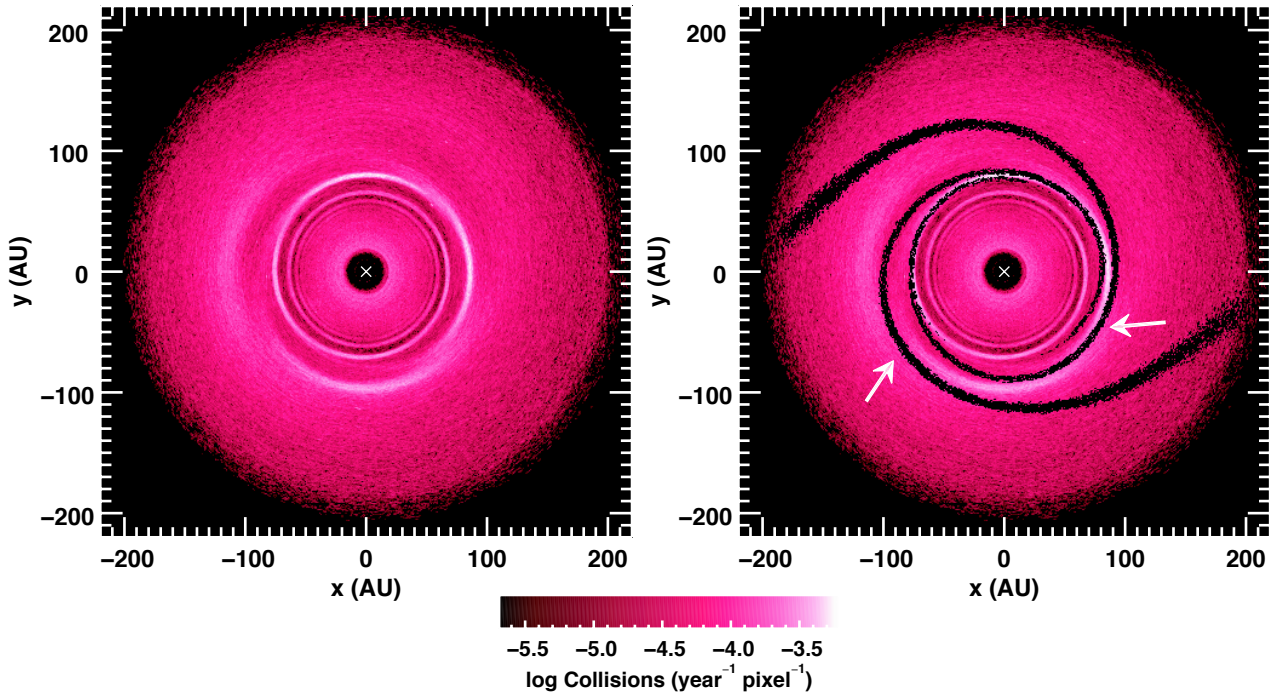}
	\caption{Left panel: Face-on map of superparticle collision rate between 10 and 10.5 Myr. The collision rate map contains a broken spiral structure corresponding roughly to the spiral density wave in Figure \ref{fig:faceon} (the ``eccentricity spiral''). Right panel: The same map, with the collisions in the planet's orbital plane masked out. The mask creates a two-armed black spiral pattern (the ``inclination spiral'') corresponding to the red and blue x's in Figure \ref{fig:orbits}. The collision rate drops where the eccentricity spiral meets the inclination spiral, creating the complex azimuthal structure. Two such breaks in the collision rate spiral are shown by the white arrows.}
	\label{fig:collboth}
\end{figure*}

SMACK allows us to model the 3D spatial distribution of collisions between parent bodies in the disk and to simulate spatial maps of dust production in the disk. For example, the left panel of Figure \ref{fig:collboth} shows a face-on map of the distribution of collisions in the $\beta$ Pic disk between 10 and 10.5 Myr. A spiral structure is evident in the collision rate, roughly corresponding to the spiral seen in the simulated disk image in Figure \ref{fig:faceon}. However, the spiral in the collision distribution in Figure \ref{fig:collboth} exhibits several ``breaks'' in its azimuthal structure that are not seen in Figure \ref{fig:faceon}.

These breaks can be understood by examining the 3D structure of the disk, specifically the interaction between the eccentricity spiral (the spiral density wave induced by the planet's eccentricity) and the  inclination spiral (the vertical displacement wave induced by the planet's inclination). In the edge-on image of the simulated disk in Figure \ref{fig:edgeon}, the inclination spiral appears as a vertical oscillation of the planetesimals with radius from the star. The inclination spiral also appears in Figure \ref{fig:orbits}, where blue and red x's mark the ascending and descending nodes for each orbit. The orbital nodes form a double-armed spiral, extending out to $\sim 100$ AU. As Figure \ref{fig:orbits} shows, this inclination spiral intersects with the eccentricity spiral as several azimuthal locations.

What happens where these two kinds of spiral intersect? In the right panel of Figure \ref{fig:collboth}, we plot the same face-on collision map shown in the left panel, but mask out the pixels within 0.2 AU of the plane of the planet's orbit at radii $>75$ AU. The right panel shows a two-armed spiral, like the pattern the ascending and descending nodes illustrated in Figure \ref{fig:orbits}. By comparison with the left panel of Figure \ref{fig:collboth}, we can see that deficits in the collision rate occur where the inclination spirals intersect the eccentricity spiral in the plane of the planet's orbit. Two such locations are indicated with right arrows in the right panel of Figure \ref{fig:collboth}, while at least two additional, unmarked intersection points are also visible further in.

\begin{figure*} [!ht]
	\centering
	\includegraphics[trim=0 50 0 30,clip,width=0.98\textwidth]{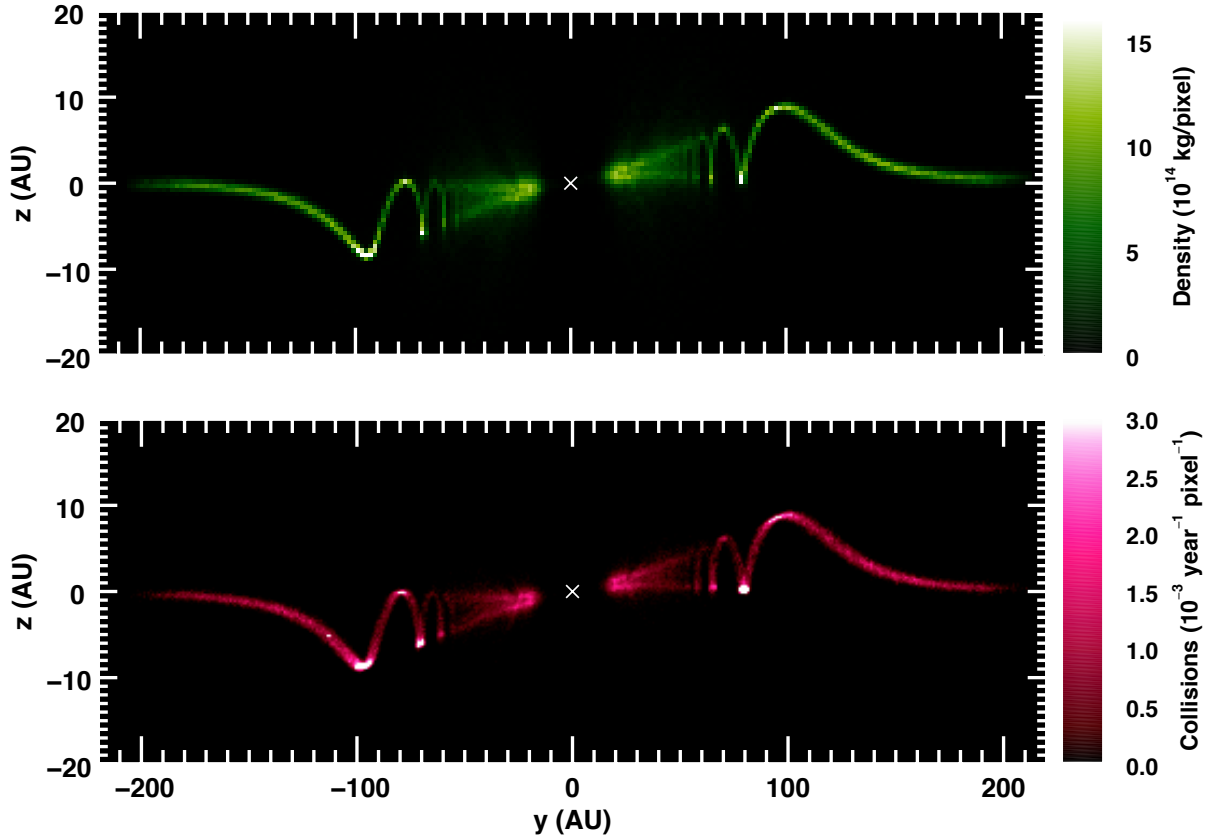}
	\caption{Mass density (upper panel) and collision rate (lower panel) of the simulated $\beta$ Pic disk at 10 Myr, cut through the $x=0$ plane. The density and collision rate are enhanced in the crests and troughs of the secular wave.}
	\label{fig:densitycut}
\end{figure*}

One reason the collision rate drops in the plane of the planet's orbit is that the density drops in this plane -- not the surface density, but the mass density. This density drop is shown in Figure \ref{fig:densitycut}, where we plot a cut through the mass density of the simulated $\beta$ Pic disk at 10 Myr in the the $x=0$ plane. We also show a cut through the collision rate in the $x=0$ plane. The density and collision rate are enhanced at the vertical peaks of the inclination wave and the troughs in midplane of the initial planetesimals disk, and minimized where the planetesimals cross the plane of the planet's orbit. When the disk is viewed face-on (as in Figure \ref{fig:faceon}), projection effects mask this effect, and the spiral density wave appears continuous. 

ALMA observations show an azimuthally-asymmetric distribution of short-lived CO gas at $\sim 85$ AU \citep{Dent2014}. The deprojected distribution of this gas appears as either two clumps of gas orbiting on either side of the star, or a single clump with a long tail. Two major hypotheses have been suggested for the origin of the ALMA asymmetry: dust production from larger planetesimals trapped in a resonant orbit by a hypothetical second planet, producing the two-clump distribution, or the recent breakup of a massive body, corresponding to the single-clump distribution \citep{Telesco2005, Dent2014}. This second scenario was modeled by \citet{Jackson2014}, who found that the timescale of the observable signature of such a breakup could be as long as 1 Myr, but that the resulting clump would be stationary in the system. 

\citet{Dent2014} concluded that the CO in the disk must be replenished with a steady-state production rate of $\sim1.4\times10^{18}$ kg/yr. Assuming a CO ice release fraction of 0.1, they estimated that the mass of solid bodies experiencing collisions must be $\sim1.4\times10^{19}$ kg/yr to maintain this CO production rate. The total mass of bodies involved in catastrophic collisions in our SMACK simulations at 10 Myr is only $1.2\times10^{16}$ kg/yr. However, our simulations only track dust-producing parent bodies up to 10 cm in diameter. Extrapolating a simple power-law size distribution with index $-3.5$, we estimate that extending our initial size distribution to include bodies  up to at least 158 km in diameter would produce the required $1.4\times10^{19}$ kg/yr of colliding mass in the simulated disk. This would provide enough colliding mass in the entire disk to produce the observed CO, but note that the CO in the $\beta$ Pictoris disk is concentrated in one (or two) smaller subvolumes of the disk. Including still larger bodies may be required to produce the observed level of CO. Future simulations are needed to test this hypothesis. \citet{Czechowski2007} described the production of gas in the $\beta$ Pic disk via high-velocity collisions, based on a simple 2D empirical description of the collision rate. We hope to improve upon this model in a future paper with SMACK, which can simulate both the 3D distribution of collisions and the collisional velocities in the $\beta$ Pic disk.

\subsection{Size Distributions vs. Disk Radius}
\label{sec:sizedists}

We also examined the size distribution of the parent bodies as they experienced collisions in SMACK. First, we calculated the average size distributions of superparticles orbiting at radii of 50, 100, 150, and 200 AU between 10.00 and 10.01 Myr, with radius bin widths of 10 AU. We fit a power law to each distribution and found that the indices varied by less than 0.3 from the initial index of $-2.5$. This relatively consistent slope indicates that the planetesimal size distribution varies only slightly with radius, so the radial mass distribution of the simulated disk probably serves as a reasonable proxy for the radial optical depth distribution at submm wavelengths.

\begin{figure} [!ht]
	\centering
	\includegraphics[scale=0.3]{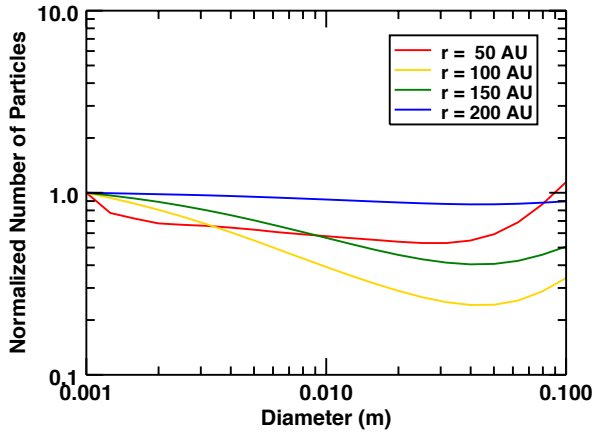}
	\caption{Average size distributions of the superparticles at various radii in the simulated disk at 10 Myr, divided by a power law with index -2.5 and normalized such that the number of particles in the smallest bin is 1 for each distribution. The size distribution becomes steeper as the secular wave from the planet approaches, then flattens back towards the initial power law.}
	\label{fig:flatdists}
\end{figure}

However, these slight variations in the size distribution may provide information about the effects of the secular perturbations on the local collision rate. In Figure \ref{fig:flatdists}, we plot each size distribution for comparison. We have divided the distributions by a power law with index -2.5 to enhance the differences between the distributions and normalized such that the each distribution equals 1 for the smallest size bin. Figure \ref{fig:flatdists} apparently shows that the spiral density wave induced by the secular perturbations from the planet shifts the size distribution away from the classical power law of index -2.5. At 10 Myr, the secular wave has not yet reached 200 AU, and the size distribution is still close to the initial power law. However, at the leading edge of the wave at 150 AU, the size distribution becomes steeper. At 100 AU, near the peak density of the spiral density wave, the size distribution is even steeper. But after the wave has passed, at 50 AU, the size distributions flattens back towards the -2.5 power law, with slight deviations at the large and small ends. Future collisional models should investigate this intriguing phenomenon. 

\section{Integrating the Dust Orbits}
\label{sec:dust}

The SMACK models described above predict where dust grains are generated in the $\beta$ Pictoris disk and their initial orbits. To understand the distribution of this dust and to model images of the disk in scattered light, we fed the output of SMACK into a second N-body integrator to track the dust orbits. Then, using a popular technique \citep[e.g.,][]{Dermott1999, Liou1999, Wilner2002, Moran2004, Stark2008, Debes2009, Kuchner2010}, we recorded the output of the integrator in a 3D histogram to simulate the density distribution of the dust cloud.

We first generated a list of all the mass production events that took place in the simulation between 10.00 and 10.01 Myr.  Every superparticle collision found by SMACK yields two such mass production events, generally producing different amounts of mass with different initial orbits. We recorded the semi-major axis, eccentricity, inclination, longitude of ascending node, argument of pericenter, mean anomaly, and total mass of dust produced for each event. The mass produced represents the mass of the fragments between 1 $\mu$m and 1 mm, distributed in a power law with an index of $-0.93$ for incremental logarithmic bins (corresponding to a size distribution power law with index $-2.8$). Note that this mass only represents the mass of dust produced in the SMACK simulation during a 10,000 yr period, so the results we present only illustrate a relative spatial distribution rather than an absolute mass of dust. See Section 8 for further discussion of this point.

\begin{figure*} [!ht]
	\centering
	\includegraphics[scale=0.4]{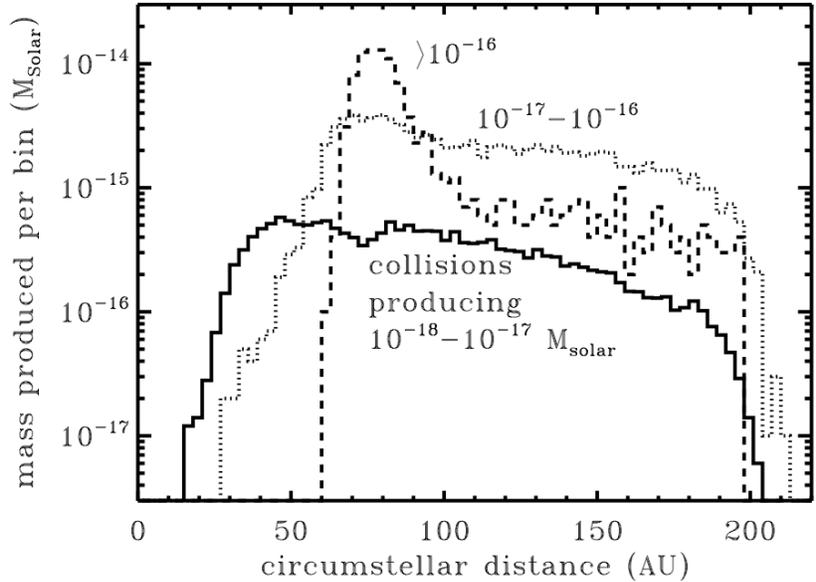}
	\caption{Radial distribution of the mass of dust produced by collisions in the SMACK simulation, binned according to mass. The most massive dust production events are confined mostly to a ring between 59-100 AU, while other collisions are spread radially through the disk.}
	\label{fig:dustmass}
\end{figure*}

Figure \ref{fig:dustmass} illustrates the radial distribution of the most significant mass production events, binned according to mass production into bins containing collisions with mass production from $10^{-18}$ to $>10^{-17}$, $10^{-17}$ to $10^{-16}$ and $>10^{-16}$ solar masses. This figure reveals an interesting result. The biggest mass production events are mostly located in a ring roughly from 59 AU to 100 AU from the star, but other collisions are spread over a wider range of circumstellar distance, and even dominate the mass production at some radii. Our results indicate that only $46\%$ of the dust, by mass, is produced in the planetesimal belt (60-90 AU). In other words, attempting to apply the ``birth ring'' approximation to describe the $\beta$ Pictoris disk \citep{Strubbe2006} would miss major sources of dust outside this ring. 

As a next step, we chose a subset of these mass production events to feed to the second N-body integrator. We selected all 960 mass production events creating $>10^{-16}$ solar masses of dust. We selected a similar number of mass production events in the range $10^{-18}$ to $>10^{-16}$ solar masses of dust produced, by choosing at random $3.5\%$ of all the events in this range. 

It is thought that the grains that dominate images of collision dominated disks like the $\beta$ Pictoris disk are those near the blowout size, i.e. with $\beta \sim 1/2$ where $\beta$ is the magnitude of the force on the grain from radiation pressure divided by the magnitude of the force on the grain from stellar gravity \citep{Strubbe2006}. We therefore sampled 6 different values of $\beta$, logarithmically distributed, near $\beta=1/2$. These values are listed in Table \ref{tab:dust}. The grain radius corresponding to a given $\beta$ value depends on assumptions about the grain's shape, composition, and optical parameters. To illustrate the range of possibilities, we list three different cases in Table \ref{tab:dust}: the simple geometric case with solid spherical grains with $0\%$ porosity, and two scenarios modeled by \citet{Augereau2001}: $4\%$ ice with $98\%$ porosity and $10\%$ ice with $95\%$ porosity. For each case, we assumed a density of 3 g/$\mbox{cm}^3$ for the rock component. 

Table \ref{tab:dust} lists values for the corresponding grain radii under three possible assumptions about the grain properties, as well as the fraction of the total dust surface area represented by each radius bin. In total, we integrated the orbits of 11,520 grains, assuming the simplified case of solid spherical grains. Since the dynamics of grains with $\beta<0.1$ are similar, we weighted the $\beta=0.09$ grains more to represent all grains up to a size of 1 mm.

\begin{deluxetable*}{l | c c c | c c c | c c c}
\tablewidth{0pt}
\tablecaption{The $\beta$ values sampled for our dust orbit integrations \label{tab:dust}}
\tablehead{\colhead{} & \multicolumn{3}{c}{(a)}  & \multicolumn{3}{c}{(b)} & \multicolumn{3}{c}{(c)} 	}
\startdata
	$\beta$	& $s$	& $m$ 	& $f_A$	& $s$	& $m$ 	& $f_A$	& $s$	& $m$	& $f_A$	\\
	0.09		& 21		& 14		& 0.26	& 260	& 4400 	& 0.42	& 110 	& 740 	& 0.38	\\
	0.14		& 13 		& 3.6 	& 0.057	& 160 	& 1100	& 0.045	& 67 		& 190  	& 0.048	\\
	0.23		& 8.3 	& 0.90  	& 0.082	& 100 	& 280	& 0.065	& 42 		& 47		& 0.069	\\
	0.36		& 5.2 	& 0.23 	& 0.12	& 65 		& 69		& 0.093	& 27 		& 12   	& 0.099 	\\
	0.57		& 3.3		& 0.057 	& 0.17	& 41 		& 17		& 0.13	& 17  	& 3.0		& 0.14 	\\
	0.90		& 2.1		& 0.014	& 0.25	& 26 		& 4.4 	& 0.19	& 11		& 0.74 	& 0.21 	
\enddata
\tablecomments{We list the corresponding radii ($s$, in $\mu$m), masses ($m$, in $10^{-9}$ g), and the fraction of the total dust surface area represented by each bin ($f_A$) for three different cases: (a) Geometric optics, solid spherical grains, $0\%$ porosity, (b) \citet{Augereau2001} case for $4\%$ ice, $98\%$ porosity, (c) \citet{Augereau2001} case for $10\%$ ice, $95\%$ porosity. Note that the $\beta=0.09$ bin includes grains up to 1 mm in diameter, resulting in much larger values of $f_A$.}
\end{deluxetable*}

Our numerical integrator uses a standard Bulirsch-Stoer algorithm \citep[see e.g.,][]{Press2007}, modified to include terms for radiation pressure and Poynting-Robertson drag \citep{Burns1979, Wilner2002, Moran2004}. It integrates the equation of motion for a dust grain \citep{Robertson1937},
\begin{equation} \frac{d^2{\bf r}}{dt^2} = -\frac{GM_{\star}(1-\beta)}{r^3}{\bf r}-\frac{GM_{\star}}{r^3}\frac{\beta}{c}[{\dot{r}}{\bf r} + r{\bf v}], \end{equation}
where $\bf r$ and $\bf v$ are the circumstellar position and velocity vectors of the grain. To the right-hand side of this equation, we added the gravitational force from the orbiting planet, though we saw no evidence of planet-dust interactions (see Section \ref{sec:dusttime}), probably because the planet is located so far interior to most of the dust. We chose a time step of one year and set the integrator to output the positions of the dust grains every $\sim211.56$ yr (i.e., 10.25 planet orbits). When the grains are created, their initial orbits conserve their birth velocities, resulting in high initial eccentricities, $e \approx \beta/(1-\beta)$, for grains created in collisions between two bodies on low eccentricity orbits. As expected \citep{Wyatt2005b, Strubbe2006}, we saw little Poynting-Robertson evolution of the dust grain orbits during the simulation of this collision-dominated disk, except for the largest values of $\beta$. Since $\beta$ Pictoris is an A-type star, we neglected the effects of stellar winds.

We ran the integration for the grain-grain collision time \citep{Wyatt1999a}, under the simplifying assumption that grains of this size are either vaporized during a collision or broken into daughter grains that contribute negligibly to the optical depth because they are quickly removed by radiation pressure. We approximated the local grain-grain collision time as a function of circumstellar distance by calculating the local optical depth of the material tracked by the superparticles, and extrapolating down 1 $\mu$m to approximate the contribution of smaller grains to the local optical depth.  We collected the output coordinates into three histograms matching the bins used for the SMACK simulated images: one for the face-on image with 2 AU by 2 AU bins, and two for two orthogonal views of the disk edge-on, with 2 AU by 0.5 AU bins. These histograms represent a steady-state system in which the dust is continuously replenished. We summed together the histograms representing dust from the biggest mass production events with the histograms representing the  dust from the mass production events in the range $10^{-18}$ to $>10^{-16}$ solar masses, weighting the latter histograms by 1.0/0.035 to compensate for our sparse sampling of these latter mass production events. We also applied the weightings $f_A$ listed in case (a) shown in Table \ref{tab:dust}.

\section{Simulated Scattered Light Morphology}
\label{sec:scatteredlight}

\begin{figure*} [!ht]
	\centering
	\includegraphics[scale=0.4]{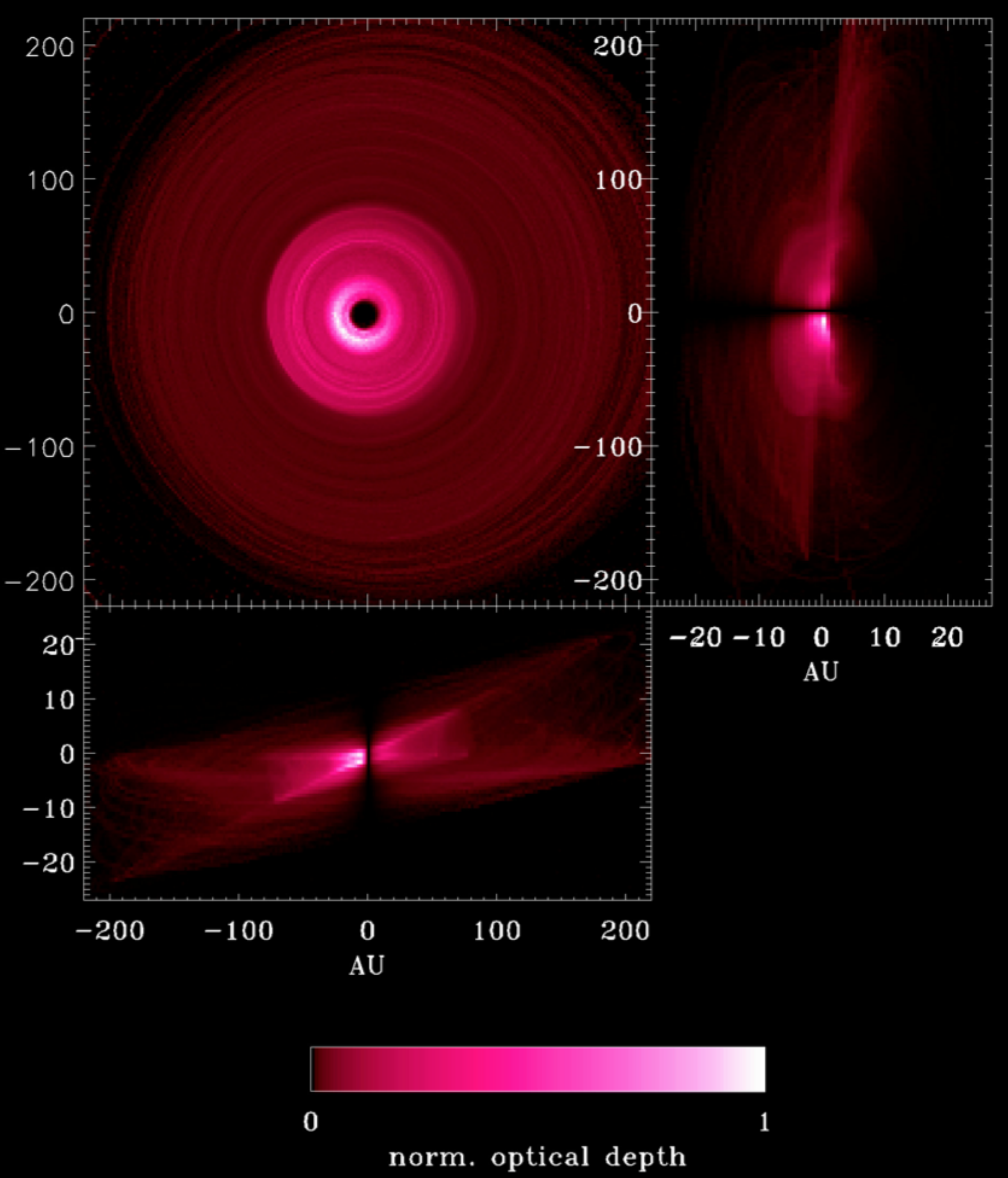}
	\caption{Simulated images of the $\beta$ Pictoris disk in scattered light, multiplied by the distance to the star squared. The dust does not show the same deficit in the inner region of the disk as the mm-sized grains. The edge-on view in the bottom panel reveals the two-disk ``x''-pattern seen in HST images.}
	\label{fig:dust3panel}
\end{figure*}

To compare our simulated dust distributions with images of the disk in scattered light, we synthesized images from our dust density histograms assuming that the dust is illuminated by 0.5 $\mu$m light from $\beta$ Pictoris, and scatters the light via a Henyey-Greenstein scattering phase function. \citet{Stark2014} found that the disk HD 181327 was not well-fit by a single Henyey-Greenstein scattering phase function. Likewise, \citet{Ahmic2009} modeled images the $\beta$ Pictoris disk taken by the Advanced Camera for Surveys (ACS) on the Hubble Space Telescope \citep{Golimowski2006} using a model consisting of a pair of thin intersecting disks, with two different values of the Henyey-Greenstein parameter, $g$. In our physical model, which is not meant to be a true inverse model like \citet{Ahmic2009}, there is no a priori physical distinction between the dust in one ``disk'' or another, so we used a single value for all the dust, the mean of the values in \citet{Ahmic2009}: $g=0.743$. Figure \ref{fig:dust3panel} shows the three simulated images, showing the normalized histograms of the optical depth of the dust from three different viewing orientations.  Each image has been multiplied by the circumstellar distance squared to highlight the faint features toward the outer edge of the disk by revealing the underlying particle distribution. 

The face-on distribution of the dust grains exhibits a spiral structure, reminiscent of the spiral structure in the planetesimal density distribution (Figure \ref{fig:faceon}). However, unlike in the planetesimal distribution, which has a deficit of larger grains in the inner region of the disk ($\lesssim59$ AU), there is an enhancement of dust grains in the inner region. This enhancement does not represent material falling inwards towards the star. It arises from the enhanced collision rate in the $\lesssim59$ AU region of the disk. The planetesimals remaining in this region collide violently and often, due to the phase mixing of their ascending nodes and pericenters. While our dust model does not simulate the 3D collisional destruction of dust in the disk, our simulations are supported by a spectroscopic detection of silicate grains inwards to 6 AU in the $\beta$ Pic disk by \citet{Okamoto2004}.

By comparing the bottom panel of Figure \ref{fig:dust3panel} with the edge-on simulated image of the planetesimals in Figure \ref{fig:edgeon}, we note that the dust grains on inclined orbits that make up the warp extend to greater circumstellar distances than the planetesimals as they are pushed outwards by radiation pressure. Figure \ref{fig:dust3panel} shows that while a significant amount of the dust orbiting inclined to the main disk is within the radial extent of the planetesimal warp (i.e., $\lesssim95$ AU), there are inclined dust grains orbiting out to $\sim200$ AU. 

\begin{figure*} [!ht]
	\centering
	\includegraphics[scale=0.6]{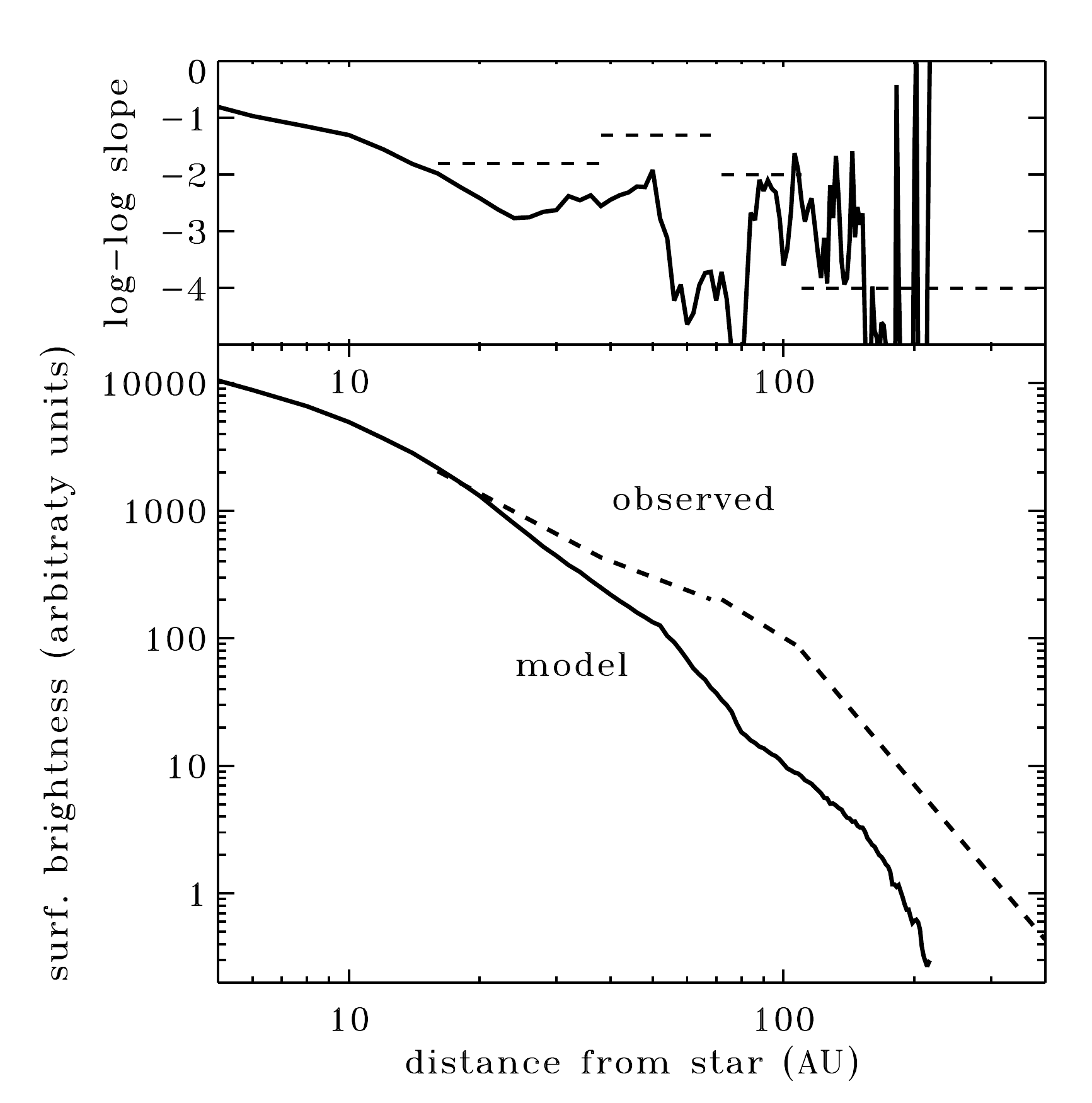}
	\caption{Radial surface brightness profile of our dust model (solid black line) compared with the power-law fits (dashed line) of \citet{Apai2015}. The grey line shows the surface brightness profile of our model. The upper panel shows the slope of the radial surface brightness profile for our models compared with the \citet{Apai2015} fits.}
	\label{fig:dustprofile}
\end{figure*}

Figure \ref{fig:dustprofile} offers a more quantitative comparison of the models to scattered-light observations of the $\beta$ Pictoris disk.  It shows an edge-on view of the model seen in the bottom panel of Figure \ref{fig:dust3panel} (solid black curve) compared to the surface brightness of the disk measured by the ACS with arbitrary brightness scaling. The observational data (dashed curve) are the power law fits to the radial surface brightness profile listed in Table 2 of \citet{Apai2015}. Models and data represent the vertically-averaged surface brightness, and the models depict the side of the disk that is along the direction of the planet's apocenter. The top panel of Figure \ref{fig:dustprofile} shows the slope of the model vs. radius from the star, compared with the power law indices fit by \citet{Apai2015}. Different observational papers cited in \citet{Apai2015} used different vertical heights to derive their brightness profiles, but their slopes agree to a level of 0.1, typically, in each of the four regions. For our model, we averaged over the vertical height of $\pm20$ AU.

Our simulated brightness profile appears to differ significantly from the observed scattered light profiles show in Figure \ref{fig:dustprofile}, revealing the limitations of our simple dust model. For example, our dust model appears to be missing a significant dust component. Figure \ref{fig:dustprofile} shows a deficit in surface brightness in our model past $\sim20$ AU, and a glance at the ACS images of the $\beta$ Pictoris disk shows that our models contain too little dust in the disk midplane. One possible reason for these mismatches is that our SMACK models used the wrong initial radial distribution of planetesimals. For example, to conserve computing time, we only included planetesimals out to a circumstellar distance of 200 AU. See Section \ref{sec:limitations} for a more detailed discussion of the limitations of our dust model.

Our simulated surface brightness profile in Figure \ref{fig:dustprofile} exhibits breaks in the the radial surface brightness power law, similar to but located radially inwards from the observed breaks. The simulated brightness profile would need to be scaled radially by a factor of at least 1.35 to fit the gross morphology of the observed profile. Given that the radial structure in our simulated disk propagates outwards in time, Equation \ref{eq:speed} suggests three possible explanations for this discrepancy: either (a) the age of the planet's orbit may be longer than 10 Myr, (b) the mass of $\beta$ Pic b may be larger than our simulation value of 9 $M_{Jup}$, and/or (c) the semimajor axis of $\beta$ Pic b's orbit may be greater than our simulation value of 9.1 AU. Considering each of these scenarios individually, Equation \ref{eq:speed} requires an age of 28 Myr, a planet mass of 25 $M_{Jup}$, or a planet semimajor axis of 15 AU to scale the radial structure of the simulated disk by a factor of 1.35. These values are clearly ruled out by observations of the planet's mass and orbit and the age of the system.

A roadblock to interpretations of the radial structure of the $\beta$ Pic disk is the contradiction between the observed radial extents of the vertical waves in the dust and planetesimal populations. For example, \citet{Wilner2002} measured the radius of the planetesimal belt to be $\sim94$ AU but \citet{Dent2014} measured the radius of the planetesimal belt as $\sim60$ AU, while scattered light observations \citep{Heap2000, Golimowski2006} measured the radial extent of the warp in the dust to be $\sim85$ AU. Future models should attempt to reconcile these differences to place constraints on the age of the planet's current orbit.

\subsection{No Structures Orbiting With The Planet's Mean Motion}
\label{sec:dusttime}

By choosing an output time step for the integrator of precisely 10.25 planet orbits, we can easily use our models to search for disk structures that evolve on the time scale of the planet's orbit, or at harmonics of the planet's orbit \citep[see][]{Wilner2002, Moran2004}. We constructed four new ``stroboscopic'' dust density histograms for this purpose: one from coordinates output when the planet's mean anomaly was 0, one from coordinates output when the planet's mean anomaly was 0.25, and so on.  These histograms, like four frames of a movie, can reveal blobs that are otherwise smeared out in other representations.

When we constructed these histograms and differenced them to search for moving patterns, we found nothing but Poisson noise.  The histograms are subject to Poisson noise from the finite number of particles per pixel; with 11520 particles and typically 700 output steps per each stroboscopic histogram, the pixels have Poisson noise at the level of about 1-$\sigma = 7$\% per 2 AU by 2  AU pixel in the face-on view. So, given the limitations of our model, we predict that there is probably no structure rotating with planet's mean motion that is stronger than about 21\%, even at the planet's semi-major axis.

This result stands in contrast to \citet{Apai2015}, who reported a marginal detection of a difference between images of $\beta$ Pictoris taken with the Hubble Space Telescope/Space Telescope Imaging Spectrograph in 1997 \citep{Heap2000} and images taken with the same instrument in 2012.  \citet{Apai2015} interpret this difference, seen at $3\arcsec-6\arcsec$ (58-117 AU) as a possible 50\% perturbation to the surface density. Future observations of this system at smaller inner working angles and future dynamical models that focus more on the inner region of the disk should resolve this issue.

\section{Limitations of the Models}
\label{sec:limitations}

We have used our physical numerical models to explore and flesh out the prevailing picture of the $\beta$ Pictoris disk as a disk of planetesimals and dust sculpted by the single planet on an inclined orbit. But before we continue our discussion of how the models compare to the observations, it is important to note the limitations of our physical models.

One of the major limitations of this work is the use of two separate models: the SMACK simulations to model the dynamics and dust-producing collisions of the parent bodies $>1$ mm in size, and the non-collisional dust dynamics model to trace the orbits of the dust grains under the influence of radiative forces. These two different models were applied sequentially, and simulated the evolution of the two populations, parent bodies and dust grains, with different approximations (no radiative forces for the parent bodies, and extremely limited collisional processing for the dust grains). This also prevented us from considering interactions between the populations.

SMACK provides a 3D distribution of the collision rate, revealing the complex azimuthally-asymmetric structures shown in Section \ref{sec:collisions}. However, in our SMACK simulation, we neglect planetesimals larger than 10 cm as a source of mass in the collisional cascade. Thus, while we have chosen the initial conditions for our SMACK models to roughly match the current optical depth inferred from optical images, we may not have correctly modeled how the absolute flux in the disk evolved prior to the present day. Our SMACK models also neglect cratering collisions, another significant source of dust \citep{Thebault2007, Muller2010, Krivov2010}. Future simulations that attempt to model the absolute flux of the dust in $\beta$ Pic should include the larger planetesimals as a source of mass as well as cratering collisions.  Adding second-order effects such as more accurate grain structure and composition to future models may also improve the accuracy of the simulated dust production in the disk.

Our dust models assume that collisions between small dust grains ($<1$ mm) are completely destructive, and calculates these collisions simply by removing grains after their collisional lifetime has expired. For the smallest ($\sim1$ $\mu$m) dust grains, the collisional products would be smaller than the blowout size and immediately removed from the system, but this assumption ignores the effects of fragmenting collisions of larger ($\sim100$ $\mu$m) grains, whose fragments may survive but whose collisions are not tracked by SMACK or our dust model. We therefore cannot accurately model the size distribution of the dust grains. More advanced dust simulations such as a suitably modified collisional grooming algorithm \citep{Stark2009} (which does not currently include dust grain fragmentation), or the dust collision algorithms of \citet{Kral2013} or \citet{Vitense2014} (which treat dust grain fragmentation more thoroughly), could be used to capture the complex evolution of the dust population.

\section{Discussion and Summary}
\label{sec:conclusions}

We have developed the first dynamical model for the $\beta$ Pictoris disk combining the colliding planetesimals, the dynamics of the resulting dust grains and the best-fit orbital parameters for the planet $\beta$ Pic b. Here, we discuss several features observed of the $\beta$ Pic disk and consider the possible origins of these features by comparing observations to our SMACK simulations and analysis.

\begin{enumerate}

\item Using SMACK's ability to model the dynamical effects of collisions, we showed that the free inclinations of the planetesimals are not damped significantly by collisions in the age of the system (Figures \ref{fig:hkdiagram} and \ref{fig:ivsa}). Our SMACK simulations reproduced the warp seen in both submillimeter \citep{Dent2014} and scattered light observations \citep[][etc.]{Burrows1995, Heap2000, Golimowski2006} of the disk, confirming that this structure can be induced by the planet's inclination, in agreement with numerical simulations by \citet{Mouillet1997}, \citet{Augereau2001}, and \citet{Dawson2011}. 

\item We showed that the spiral density wave in the mm-sized planetesimals induced by the planet's eccentricity extends out to roughly the same radial extent as the warp (Figures \ref{fig:edgeon} and \ref{fig:faceon}), and could be interpreted as a ring or belt in edge-on observations \citep{Wilner2011, Dent2014}.

\item Our SMACK simulations also demonstrated that the deficit in mm-sized grains interior to this belt \citep{Wilner2011, Dent2014} is created via the mechanism described by \citet{Mustill2009}: collisions are excited by the planet's eccentricity interior to the spiral density wave, eroding and removing material from the inner region of the disk and creating an observable deficit (Figure \ref{fig:radmoid}). 

\item We propose a new mechanism for producing azimuthally-asymmetric dust  and gas \citep{Wahhaj2003, Telesco2005, Li2012, Dent2014} via collisions, without invoking MMRs or massive collisions \citep{Telesco2005, Dent2014, Jackson2014}: an azimuthally-asymmetric collision rate along the spiral density wave (Figure \ref{fig:collboth}), created by the interaction between the spiral density wave (induced by the planet's eccentricity) and the vertical displacement wave (induced by the planet's inclination). Though the collisions among the bodies we modeled ($<10$ cm in size) do not themselves produce enough CO to match the ALMA observations, collisions among slightly larger parent bodies in the same disk structure probably could.

\item We predict that there are no dust structures orbiting with the planet that are detectable above the Poisson noise of the simulation, supporting the suggestion that the $50\%$ perturbation measured by \citet{Apai2015} may have been produced by a recent massive collision, since such collisions were not included in our simulations.

\item Our SMACK simulations of the dust production in the disk revealed that only $46\%$ of the dust is produced in the planetesimal belt (60-90 AU), indicating that the ``birth ring'' approximation \citep{Strubbe2006} fails to account for over $50\%$ of the mass of dust produced via collision in the disk. Instead of a ``birth ring'' at this location, the $\beta$ Pictoris system appears to have a ``stirring ring'' where only the high-velocity planetesimal collisions are concentrated (Figure \ref{fig:dustprofile}).

\item Our simulated dust distribution (Figure \ref{fig:dust3panel}) reproduced the x-shaped pattern seen in high-resolution scattered light images of the edge-on disk by \citet{Golimowski2006, Ahmic2009, Apai2015}. This pattern arises because more dust is produced in the peaks and troughs of the secular wave (Figure \ref{fig:densitycut}). It has not been reproduced by previous dynamical models of the dust \citep{Mouillet1997, Augereau2001, Dawson2011}.

\end{enumerate}

Though our model appears to have explained several salient features of the $\beta$ Pictoris disk, many open questions remain about the physics of this complicated planetary system. 

\textit{When and how was the planet scattered into its current orbit?} Possible mechanisms suggested by \citet{Dawson2011} include scattering by a second planet. The timescale for this scattering could be estimated by modeling the speed at which the secular perturbations of the planet propagate outwards through the disk. However, the observational evidence for the radial extent of the secular perturbations appears contradictory. \citet{Wilner2002} and \citet{Dent2014} measured brightness peaks in the submillimeter observations, which they interpreted as the planetesimal belt. But \citet{Wilner2002} placed this belt at a radius of $\sim94$ AU while \citet{Dent2014} measured it at $\sim60$ AU. Scattered light observations of the disk indicate that the warp in the dust distribution extends out to $\sim85$ AU \citep{Heap2000, Golimowski2006}. An inverse model that includes the various measurements of the warp and stirring belt could use the radial extent of these secular perturbations from the planet to predict when the planet was scattered onto its current orbit.

\textit{What is the role of gas in shaping the dust in the disk?} \citet{Thebault2005} argued that the long-term effects of gas drag alone probably have had a negligible effect on the dust in the $\beta$ Pictoris disk. However, \citet{Lyra2013} demonstrated that instabilities involving gas drag, photoelectric heating and streaming effects can cause clumps and rings to form in debris disks like $\beta$ Pictoris. 

\textit{What is the role of large planetesimals in shaping the disk?}  Our simulations did not track bodies larger than 10 cm in size, but the rare collisions between these massive bodies could yield important contributions to the observable debris \citep[e.g.,][]{Jackson2014,Stark2014, Kral2015}.

\textit{What is the origin of the NE-SW brightness asymmetry?} The difference in surface brightness between the NE and SW wings of the disk has been observed at various wavelengths, including the optical, infrared, and submillimeter \citep{Apai2015}. But perhaps the planet's orbit may be more eccentric than current measurements indicate, or perhaps there are additional planets in the system creating this asymmetry.

The 3D planetesimal and dust model we used for the $\beta$ Pic disk could also be applied to other interesting debris disk systems. AU Microscopii, for example, is another edge-on disk with a planetesimal belt observed in the submillimeter \citep{MacGregor2013} and small-scale structure in the dust disk \citep{Fitzgerald2007}. A combined planetesimal and dust model are needed to connect the submillimeter and scattered light observations, by understanding how the dynamics of the planetesimals in the presence of a hypothetical planet affect the locations of dust production events. Previous modelers have also applied the ``birth ring'' approximation to the AU Mic disk \citep{Fitzgerald2007, MacGregor2013}, but we have shown that the birth ring approximation can miss over half of the sources of dust in a disk with a planetesimal belt, indicating that observations of AU Mic should be revisited with a combined planetesimal and dust model to account for these sources.

The HD 15115 debris disk, often referred to as the ``blue needle'' or more recently, the ``grey needle'' \citep{Rodigas2012}, exhibits a brightness asymmetry despite its morphological symmetry, and was recently found to have a large central cavity \citep{Mazoyer2014}. Collisional clearing may be responsible for the central clearing, and the asymmetric collision effect could create a brightness asymmetric in the absence of an offset between the disk and star. The edge-on HD 32297 also exhibits a brightness asymmetry and a central clearing \citep{Currie2012a}, and should be studied further with a collisional model.

The asymmetric collision effect may also be observable in disks that are not viewed edge-on. For example, high-resolution near-IR imaging of the HD 141569A disk has revealed an inner spiral feature inclined to the outer ring, and a cleared inner region \citep{Biller2015}. A planet on an inclined and eccentric orbit could be responsible for such a disk morphology, and may also produce observable dust clumps via to asymmetric collisions. High-resolution submillimeter observations of HD 141569A (for example, with ALMA) are needed to search for azimuthally-asymmetric gas distributions, and simulations like the SMACK and dust models described in this work could constrain the mass and orbit of a possible exoplanet orbiting in the disk.

\vspace{1em}

Erika Nesvold and Marc Kuchner are supported in part by NASA Planetary Geology and Geophysics grant PGG11-0032. Erika Nesvold is supported in part by the ALMA Student Observing Support Program through NRAO. Marc Kuchner is supported in part by the NASA Astrobiology Institute through the Goddard Center for Astrobiology.  

\bibliographystyle{./apj}
\bibliography{./betaPic}

\newpage

\end{document}